\documentclass[11pt]{article}
\usepackage[mathscr]{eucal}
\usepackage{
amsmath,amsfonts,amssymb,amsthm,bm,calrsfs,stmaryrd}
\usepackage{hyperref}

\def\bea{\begin{eqnarray}}
\def\eea{\end{eqnarray}}
\def\be{\begin{equation}}
\def\ee{\end{equation}}
\def\ba{\begin{array}}
\def\ea{\end{array}}
\def\eq#1{(\ref{#1})}

\def\yb{\bar{y}}
\def\zb{\bar{z}}

\def\cI{\mathcal{I}}

\def\bea{\begin{eqnarray}}
\def\eea{\end{eqnarray}}
\def\ba{\begin{array}}
\def\ea{\end{array}}

\def\nb{\bar{\nu}}
\def\a{\alpha}
\def\b{\beta}
\def\c{\gamma}
\def\d{\delta}
\def\e{\epsilon}

\def\s{\sigma}
\def\t{\tau}

\def\L{\Lambda}

\def\o{\omega}

\def\ad{\dot{\alpha}}
\def\bd{\dot{\beta}}

\def\una{{\underline{\alpha}}}

\def\ad{{\dot{\alpha}}}
\def\bd{{\dot{\beta}}}

\def\l{\lambda}
\def\lb{\bar{\lambda}}
\def\m{\mu}
\def\mb{\bar{\mu}}

\def\F{\widehat{\Phi}}
\def\J{\widehat{J}}
\def\A{\widehat{A}}

\usepackage{color}
\definecolor{rougef}{rgb}{0.56,0,0}
\definecolor{vertf}{rgb}{0,0.5,0}
\definecolor{bleuf}{rgb}{0,0,0.8}

\begin{document}
\renewcommand{\thefootnote}{\fnsymbol{footnote}}
\renewcommand{\theequation}{\thesection.\arabic{equation}}\csname@addtoreset\endcsname{equation}{section}

%

\vspace{7mm}

\begin{center}
{\Large\bf Higher Spin Gravity Amplitudes From Zero-form Charges}
\vspace{1.5cm}

N i c o l \`o ~~ C o l o m b o 
~~~~ and ~~~~P e r ~~ S u n d e l l\footnote{F.R.S.-FNRS Researcher with an 
Ulysse Incentive Grant for Mobility in Scientific Research}

\vspace{15mm}

\textit{Service de M\'ecanique et Gravitation\\Universit\'e de Mons --- UMONS \\20 Place du Parc\\ B-7000 Mons, Belgium}
\vspace*{1cm}

{\footnotesize \tt  nicolo.colombo@umons.ac.be, per.sundell@umons.ac.be}

\end{center}

\vspace{1.5cm}

\begin{minipage}{.90\textwidth}

\textsc{Abstract.}
We examine zero-form charges in Vasiliev's four-dimensional bosonic higher spin 
gravities.
These are classical observables given by integrals over noncommutative twistor space of adjoint combinations of the zero-form master fields, including insertions of delta functions in the deformed oscillators serving as gauge invariant regulators. 
The regularized charges admit perturbative expansions in terms of multi-linear functionals in the Weyl zero-form, which are Bose symmetric and higher spin invariant by construction, and that can be interpreted as basic building blocks for higher spin gravity amplitudes.
We compute two- and three-point functions by attaching external legs given by unfolded bulk-to-boundary propagators, and identify the result with the two- and three-current correlation functions in theories of free conformal scalars and fermions in three dimensions.
Modulo assumptions on the structure of the sub-leading corrections, and relying on the generalized Hamiltonian off-shell formulation, we are thus led to propose an expression for the free energy as a sum of suitably normalized zero-form charges

\end{minipage}

\vspace{.9cm}

\renewcommand{\thefootnote}{\arabic{footnote}}

\setcounter{footnote}{0}

\newpage

{\small \tableofcontents }

\section{Introduction}
Vasiliev's higher spin gravities \cite{Vasiliev:1990en,Vasiliev:1992av,Vasiliev:2004cp} are the only known fully nonlinear interacting models including massless gauge fields of spin greater than two.
So far,  the theory has been examined primarily at the level of its classical equations of motion and in terms of locally defined quantities.
In the four-dimensional case, however, an action principle of generalized Hamiltonian type has been given in \cite{Boulanger:2011dd} and the related geometric framework has been studied in \cite{Sezgin:2011hq,Boulanger:2012bj}.
Within this context, a method for computing amplitudes without referring to any off-shell formulation based on Fronsdal fields,
facilitated by the salient features of unfolded dynamics and twistor calculus underlying the four-dimensional models, was proposed and examined in some detail in \cite{Colombo:2010fu}.

The method is based on extracting basic building blocks for amplitudes, referred to as quasi-amplitudes, from of a special type of classical observables, known as zero-form charges, first introduced in \cite{Sezgin:2005pv}.
The zero-form charges are fully nonlinear higher spin gauge invariant functionals 
given by integrals over twistor space, rather than spacetime, of constructs formed out of curvature tensors, rather than gauge fields. 
Using suitable regularization schemes, these charges can be expanded perturbatively in different sectors of the theory, labelled by boundary conditions, thus giving rise to multi-linear functionals in curvature tensors, which are Bose symmetric and higher spin invariant by their construction. 

In \cite{Colombo:2010fu}, it was found that the perturbative expansion in terms of twistor-space plane waves contains two sets of divergencies. One set is associated to parametric homotopy integrals used to invert the de Rham differential in twistor space, and another set to integrals over noncommutative twistor space used to represent $\star$-products and traces.
The first set was regularized using a certain closed-contour prescription that is consistent with associativity and hence gauge invariance, and the second set using a naive multiplicative normalization based on factoring out an infinite twistor-space volume measured using the constant symplectic two-form in twistor space.

In this paper, we refine our previous regularization scheme in the following two ways: i) the closed contours are collapsed onto the branch cuts leading to open contours that can be evaluated using analytical continuations and the principal value prescription; and ii) the twistor space volume to be factored out is taken to be measured using the symplectic two-form defined by Vasiliev's deformed oscillator.
We implement the new scheme using twistor space plane waves, which yields regularized zero-form charges with nontrivial sub-leading corrections.
We proceed by evaluating quasi-amplitudes with two and three external legs arising in the leading and next-to leading order by saturating these with Weyl zero-forms corresponding unfolded bulk-to-boundary propagators given in the polarization spinor basis of Giombi and Yin \cite{Giombi:2011rz}.
Not too surprisingly, the resulting two- and three-point functions agree with the two-and three-current correlation functions in the three-dimensional theories of free conformal scalars and fermions.

As the zero-form charges are finite for nontrivial exact solutions, it is natural to ask whether our on-shell method can be supplemented by a suitable off-shell input as to provide a recipe for how to combined the zero-form charges into an on-shell action interpretable as a fully nonlinear and regularized free energy.
To this end, we shall assume that our amplitude calculations admit a natural generalization to any number of points and to sub-leading corrections, which assume involve coincidence limits.
We shall also assume that certain coupling functions, arising in the twistor space regularization, can be fixed by demanding a meaningful off-shell extension of the zero-form charges as deformations of the generalized Hamiltonian action.
We are then led to propose that the free energy is given by a specific sum of zero-form charges, with coefficients fixed by the aforementioned requirements off shell as well as cluster decomposition conditions on shell,
and to identify it with the generating functional of the corresponding three-dimensional conformal field theory deformed by \emph{finite} sources.

In other words, our proposal amounts to that the nontrivial interactions in the Vasiliev system, which appear in sub-leading corrections to the zero-form invariants, correspond to coincidence contributions to the $n$-point functions in the conformal theory.
In this sense, the leading orders of the zero-form invariants and the corresponding point-split correlation functions, may be considered as trivial kinematics, while the physically interesting aspects of the theory reside in the existence of a fully non-linear free energy functional. 

The paper is organized as follows: Section 2 contains a brief review of Vasiliev's equations providing technical details as well as the on-shell concepts underlying the twistor space method. The regularization scheme is spelled out in Section 3 and applied to amplitude calculations in Section 4. Section 5 describes the enhanced twistor space method, containing an off-shell component lifted from \cite{Sezgin:2011hq,Boulanger:2012bj}, meant to provide a concrete procedure for constructing the free energy. We conclude in Section 6 with a set of selected remarks on future directions and a summary of the results of this paper.  

\section{Vasiliev's Minimal Bosonic Models}

In this section we provide a brief review of Vasiliev's minimal bosonic models \cite{Vasiliev:1992av,Sezgin:2002ru}: their parent formulation in correspondence space; reductions to dual formulations in spacetime and twistor space; and finally the zero-form charges as natural observables in twistor space.

\subsection{Correspondence Space, Master Fields and Equations of Motion}

\paragraph{Correspondence Space, Symbols and Traces}

The minimal bosonic models that we shall study are formulated in terms of master fields on a correspondence space ${\cal C}$ which is a non-commutative symplectic manifold consisting of coordinate charts 
\be
{\cal C}_I~=~T^\ast{\cal X}_{I}\times {\cal Y}\times {\cal Z}
\ee 
where ${\cal X}_I$ is a commutative manifold and ${\cal Y}$ and ${\cal Z}$ are two four-dimensional symplectic manifolds.
Letting  
\be
Y^{\una}~=~(y^\a,\yb^\ad)\ ,\quad Z^{\una}~=~(z^\a,-\zb^\ad)\ ,\quad (y^\a)^\dagger~=~\yb^{\ad}\ ,\quad (z^\a)^\dagger~=~\zb^{\ad}\ ,
\ee
and $(X^M,P_M)$, respectively, denote canonical coordinates for ${\cal Y}\times {\cal Z}$ and $T^\ast{\cal X}_{I}$, with non-trivial commutation rules\footnote{We use conventions in which $\l^\a=\e^{\a\b}\l_\b$ and $\l_\a=\l^\b\e_{\b\a}$.}
\be
 [y^{\a},y^{\b}]_\star ~=~2i \e^{\a\b}~=~-[z^{\a},z^{\b}]_\star\ ,\quad [X^M,P_N]_\star ~=~i\delta^M_N\ ,
\ee
the space of arbitrary polynomials $\widehat P (X,P,Y,Z)$ is an associative algebra with binary product $\star$.
By choosing basis elements that are labelled by classical monomials, this algebra can be represented in the space of classical polynomials, whereby $\widehat{P}$ is mapped to a classical polynomial $\widehat f_{\widehat { P}}$, referred to as the symbol of $\widehat {P}$, such that
\be \widehat f_{\widehat {\cal P}_1\star\widehat{\cal P}_2}~=~\widehat f_{\widehat {\cal P}_1}\star\widehat f_{\widehat{\cal P}_2}\ ,\ee
where the $\star$-product composition on the right-hand side
can be represented using an auxiliary integral with a suitable integration kernel, which thus depends on the basis.
Non-trivial dynamics and boundary conditions in ${\cal C}$
require non-polynomial extensions preserving associativity
that can be described using regularization schemes chosen
such that preferred sets of classical observables and related 
semi-classical amplitudes have well-defined classical perturbative 
expansions, as we shall illustrate below.
As the observables should not depend on the choice of basis, 
it is natural to construct them via the trace operations 
\bea
\label{trace}
 \widehat{\rm Tr}[ \widehat {\cal O} (Y, Z)]  ~  = ~ \int \frac{d^4U}{(2\pi)^2} \int \frac{d^4V}{(2\pi)^2} \ \widehat f_{\widehat {\cal O} }(U ,  V) \  , 
\eea 
where $d^4U:=d^2u d^2\bar u$ \emph{idem} $d^4V$, and the classical integration variables $( u^{\a} , \bar u^{\ad}, v_{\a},  \bar v_{\ad} )$ may be either complex or real doublets, referred to as the real and chiral trace operations\footnote{Thus, in the real trace operation one uses the reality condition $(u^\a)^\dagger=\bar u^{\ad}$, while in the chiral trace operation the two doubles are treated as two independent real doublets. In both cases one has $( \widehat {\rm Tr}[O] )^{\dagger}  =  ( \widehat {\rm Tr}[O^{\dagger}] ) $. }, respectively.
When finite, both trace operations are cyclic and independent of basis\footnote{A slight alteration of the ordering prescription induces a change in $\widehat f_{\widehat{\cal O}}$ given by a total derivatives on ${\cal Y}\times {\cal Z}$.} up to boundary terms on ${\cal Y}\times {\cal Z}$. 
Requiring all physical observables to be given by traces over ${\cal Y}\times {\cal Z}$ yields a theory that is invariant under similarity transformations in their turn to be identified as the higher spin gauge transformations.
In what follows, we shall use the chiral trace operation.

\paragraph{Master Fields}

The master fields are differential forms on ${\cal C}_I$ that are annihilated by $i_{\partial/\partial Y^{\una}}$, that is, of form degree zero on ${\cal Y}$. The $\star$-product composition of such objects using $[dX^M,\widehat f]_\star\equiv 0$ \emph{idem} $dP_M$ and $dZ^{\una}$ where the graded commutator
\be \left[\widehat f,\widehat g\,\right]_\star:=\widehat f\star \widehat g-(-1)^{{\rm deg}(\widehat f){\rm deg}(\widehat g)}\widehat g \star \widehat f\ ,\ee
defined a graded associative algebra with a differential
\be \widehat d~:=~ d + q\ ,\qquad d~:=~dX^{M}\partial_{M}+ dP_M \partial^M\ ,\quad q~:=~dZ^{\una} \partial_{\una}\ ,\ee
which is compatible with the $\star$-product in the sense that
 \be
 \widehat d (\widehat f \star \widehat g)   ~=~ (\widehat d\widehat f) \star \widehat g + (  - )^{{\rm deg}(\widehat f)   } \widehat f \star (\widehat d \widehat g)\ , 
 \ee
where ${\rm deg}$ denotes the total form degree. 
In the duality-unextended models, the field content consists of a zero-form 
$\widehat \Phi$ and a one-form 
\be \widehat A~=~\widehat U+\widehat V\ ,\quad \widehat U ~ =~dX^{M} \,\widehat U_{M} + dP_M \widehat U^M\ ,\quad\widehat V~=~dz^\a \widehat V_\a +  d\zb^\ad \widehat V_\ad\ ,\ee
subject to the following kinematic constraints\footnote{The maps $\tau$ and $\pi$ do not act on classical component fields while $\dagger$ acts on these fields by means of a suitable hermitian conjugation, which can be taken to be complex conjugation in the classical theory. }:
\be
\label{minimal projection}
 \tau(\widehat A,\widehat \Phi)~=~(-\widehat A,\pi(\widehat \Phi))\ ,\quad 
 (\widehat A,\widehat \Phi)^\dagger~=~(-\widehat A,\pi(\widehat \Phi))\ ,
\ee
where the graded differential-algebra anti-automorphism 
\be \t~(y_{\a},\yb_\ad;z_\a,\zb_\ad)~:=~(iy_{\a},i\yb_\ad;-iz_\a,-i\zb_\ad)\ ,\ee
\be \widehat d\circ \tau~:=~\tau\circ\widehat d\ ,\quad \t(\widehat f\star \widehat g)~:=~(-1)^{{\rm deg}(\widehat f){\rm deg}(\widehat g)} \t(\widehat g)\star\t(\widehat f)\ ;\ee
the differential-algebra automorphism 
\be
\label{pz1}
\pi~(y_{\a},\yb_\ad;z_\a,\zb_\ad)~:=~(-y_{\a},\yb_\ad;-z_\a,\zb_\ad)\ ,\ee
\be \widehat d\circ \pi~:=~\pi\circ\widehat d\ ,\quad \pi(\widehat f\star \widehat g)~:=~\pi(\widehat f)\star\pi(\widehat g)\ ,\ee
\emph{idem} $\bar\pi(y_{\a},\yb_\ad;z_\a,\zb_\ad):=(y_{\a},-\yb_\ad;z_\a,-\zb_\ad)$; and hermitian conjugation operation is chosen such that
\be (\widehat f\star \widehat g)^\dagger~=~(-1)^{\widehat f\widehat g}\, \widehat g^{\,\dagger}\star \widehat f^{\,\dagger} \ ,\quad (\widehat d\widehat f)^\dagger~=~\widehat d(\widehat f^\dagger)\ .
\ee
The $\tau$-condition implies the weaker bosonic projection $\pi\bar\pi(\widehat A,\widehat \Phi)=(\widehat A,\widehat \Phi)$ which defines a non-minimal bosonic model whose four-dimensional space-time description is in terms of symmetric tensor gauge fields of even and odd ranks. The stronger $\tau$-condition removes all odd ranks thus leaving the minimal bosonic model consisting of only even ranks.
The kinematic constraints on $\widehat U$ defines the adjoint representation of an extended higher spin algebra containing the higher spin algebra $hs(4)$ as the subalgebra that commutes to $Z^{\una}$, that is
\be hs(4)~=~\left\{~f(Y)~:~ \tau(f)~=~f^\dagger~=~-f~\right\}\ .\ee
The kinematic constraints on $\widehat \Phi$ defines an extended twisted-adjoint representation containing the twisted-adjoint representation of $hs(4)$ as the subspace of elements that commutes to $Z^{\una}$, that is  
\be T[hs(4)]~=~\left\{~f(Y)~:~ \tau(f)~=~\pi(f)~,~~ f^\dagger~=~\bar\pi(f)~\right\}\ee
of $hs(4)$.

\paragraph{Equations of Motion}

We shall focus on (minimal) bosonic models with linear interaction function \cite{Vasiliev:1990en,Vasiliev:1990cm,Vasiliev:1990vu,Vasiliev:1992av} (see also \cite{Sezgin:2002ru,Sezgin:2003pt,Boulanger:2011dd,Sezgin:2011hq}), for which Vasiliev's equations of motion are given by the curvature constraints
\be 
\widehat F+i\widehat\Phi\star (b \widehat J+{\rm h.c.})~=~0\ ,\qquad \widehat D\,\widehat \Phi~=~0\ ,\qquad \widehat d\,\widehat J~=~0\  ,\label{eom}
\ee
\be 
\widehat F~:=~\widehat d\,\A+\A\star \A\ ,\qquad \widehat D\,\F~:=~\F+\left[\A,\F\,\right]_\pi\ ,
\ee
where the graded twisted-commutator $\left[\widehat f,\widehat g\,\right]_\pi:=\widehat f\star \widehat g-(-1)^{{\rm deg}(\widehat f){\rm deg}(\widehat g)}\widehat g \star\pi(\widehat f\,)\ $,
\be
b~\equiv~e^{i \theta } \ ,\qquad \theta~\in~[0,\pi]\ ,
\ee  
and $\widehat J $ is a globally defined holomorphic two-form on ${\cal Z}$ assumed to obey 
\be 
\tau(\widehat J\,)~=-~\widehat J\ ,\quad \left[\J, \widehat f\,\right]_\pi~=~ 0  \label{Jtwist}
\ee 
for any $\widehat f$ such that  $\pi\bar\pi(\widehat f)=\widehat f$.
One may choose 
\be 
\label{J}
\widehat J ~=~ -\frac{1}{4} dz^2~\widehat{\kappa}\ ,\quad d^2z~:=~dz^\a dz_\a\ ,
\ee
where $\widehat{\kappa}$ is the Klein operator of the complexified Heisenberg algebra generated by $(y_\a,z_\a)$ to be specified below. 
The equations of motion are compatible with $\widehat d^2\equiv 0$ and hence invariant under the Cartan gauge transformations
\be
\delta_{\widehat{ \e} } \,\widehat A ~=~ \widehat d \widehat \e ~ -~  \left[\widehat  \e, \widehat A  \right]_\star   \, \quad
\delta_{\widehat{ \e} } \,\widehat\Phi~=~ -\left[\widehat  \e,\F \right]_\pi\ .
\ee
Written out in components, the equations of motion read
\be d\widehat U+\widehat U\star \widehat U~=~0\ ,\quad d\widehat \Phi+[\widehat U,\widehat \Phi]_\pi~=~0\ ,\ee
\be d\widehat V-q\widehat U+[\widehat U,\widehat V]_\star~=~0\ ,\ee
\be q\widehat V+\widehat V\star\widehat V~=~\frac{i}4 \left(b\, dz^2  \widehat \Phi\star\widehat \kappa+\bar b \,d\zb^2 \widehat \Phi\star\widehat{\bar\kappa}\right)\ ,\quad q\widehat \Phi +[\widehat V,\widehat\Phi]_\pi~=~0\ ,\ee
that is, $\widehat U$ is flat while $\widehat V$ has a curvature proportional to $\widehat \Phi$ which is in its turn covariantly constant in both ${\cal X}$-space and ${\cal Z}$-space.

\paragraph{Lorentz Covariance}

In addition, the equations are manifestly Lorentz covariant as can be seen by introducing an \emph{a priori} independent canonical Lorentz connection $\o^{\a\b}$ via
\footnote{Under a Lorentz transformation with parameter $\L^{\a\b}$, one has $\delta_\L\o_{\a\b}=d\L_{\a\b}-2\o_{(\a}{}^\c \L_{\b)\c}$ and $\delta_\L \widehat W=-\left[\widehat \e^{(0)},\widehat W\right]_\star$,  $\delta_\L\widehat \Phi=-\left[\widehat \e^{(0)},\widehat \Phi\right]_\star$ and $\delta_\L \widehat S_\a=\L_\a{}^\b\widehat S_\b-\left[\widehat \e^{(0)},\widehat S_\a\right]_\star$ where $\widehat \e^{(0)}:=\frac1{4i} (\L^{\a\b}\widehat M^{(0)}_{\a\b}+{\rm h.c.})$.}
\be \widehat W~:=~\widehat U-\widehat K\ ,\quad \widehat K~:=~\frac{1}{4i} \left(\omega^{\a\b}\widehat M_{\a\b}+{\rm h.c.}\right)\ ,\quad \widehat M_{\a\b}~:=~ \widehat M^{(0)}_{\a\b}+\widehat S_{(\a}\star \widehat S_{\b)}\ ,\ee
where the full Lorentz generators $\widehat M_{\a\b}$ are given in terms of the internal Lorentz generators $\widehat M^{(0)}_{\a\b}:=y_{(\a}\star y_{\b)}-z_{(\a}\star z_{\b)}$ and the deformed oscillators
\be \widehat S_\a~:=~z_\a-2i\widehat V_{\a}\ ,\ee
and rewriting the equations of motion as
\be \nabla \widehat W+\widehat W\star \widehat W+\frac{1}{4i} (r^{\a\b} \widehat M_{\a\b}+{\rm h.c.})~=~0\ ,\ee
\be \nabla \widehat \Phi+\left[\widehat W,\widehat \Phi\right]_\pi~=~0\ ,\quad \nabla \widehat S_\a +\left[\widehat W,\widehat S_\a\right]_\star~=~0, \ee
\be \widehat S_\a\star \widehat \Phi +\widehat \Phi\star \pi(\widehat S_\a)~=~0\ ,\ee
\be \left[\widehat S_\a,\widehat S_\b\right]_\star + 2i\e_{\a\b}(1-b \widehat \Phi\star \widehat\kappa)~=~0\ ,\ee 
where $r^{\a\b}:=d\o^{\a\b}-\o^{\a\c}\wedge \o^{\b}{}_{\c}$ and $\nabla:=d+\omega$ with $\omega$ acting canonically, \emph{viz.} \cite{}
\be \nabla \widehat W~=~d\widehat W+\frac{1}{4i} \left[\o^{\a\b}\widehat M^{(0)}_{\a\b}+{\rm h.c.},\widehat W\right]_\star\ ,\quad \nabla \widehat \Phi~=~d\widehat \Phi+\frac{1}{4i} \left[\o^{\a\b}\widehat M^{(0)}_{\a\b}+{\rm h.c.},\widehat \Phi\right]_\star\ ,\ee\be
\nabla \widehat S_\a~=~d\widehat S_\a-\o_\a{}^{\b} \widehat S_\b +\frac{1}{4i} \left[\o^{\b\c}\widehat M^{(0)}_{\b\c}+{\rm h.c.},\widehat S_\a\right]_\star\ .\ee
Under Cartan gauge transformations, one has
\be \delta_{\widehat\e}\widehat W~=~\nabla\widehat\e+[\widehat W,\widehat\e]_\star\ ,\quad \delta_{\widehat\e}\omega^{\a\b}~=~0\ ,\quad \delta_{\widehat\e}\widehat S_\a~=~[\widehat S_\a,\widehat\e]_\star\ .\ee
The introduction of $\omega^{\a\b}$ induces a shift symmetry that can be used to set to zero the $y^{(\a}\star y^{\b)}$-component of $\widehat W$; in perturbative expansions in which $\widehat M_{\a\b}$ is given by $y_{(\a}\star y_{\b)}$ in the leading order, this choice amounts to taking $\o_{\a\b}$ proportional to the $y^{(\a}\star y^{\b)}$-component of $\frac{\partial^2}{\partial y^\a \partial y^\b}\widehat U|_{Y=Z=0}$ in the leading order.

\paragraph{Parity Invariant Type A and Type B Models}

The parity map $P$, which acts as an automorphism on the associative algebra given by the direct product of the oscillator $\star$-product algebra and the algebra of component fields, is defined by
\be
P(y^\a , \yb^{\ad}, z^\a , \zb^{\ad} )  ~=~ ( \yb^{\ad},y^\a ,  - \zb^{\ad},   - z^\a   ) \ ,\quad \widehat d P~=~P\widehat d\ ,\ee
and by assigning the master fields suitable intrinsic parities, which in general may constrain the interactions.
Thus, letting 
\be \widehat f~=~\sum_\l \widehat T^\lambda(Y,Z,dZ)  f_\lambda(X,dX)\ ,\ee
denote the expansion of a master field $\widehat f$ into component fields $f_\lambda$ in sector with basis $\widehat T^\l$ that remains invariant under the action of $P$, \emph{i.e.} $P(\widehat T^\lambda)=\widehat T^{\lambda'} P_{\lambda'}^\lambda$ where $P_{\lambda'}^\lambda$ is a matrix squaring to one, one has $P(\widehat f)=\sum_\l \widehat T^\lambda P_\lambda^{\lambda'} P( \widehat f_\lambda)$, and the action of $P$ on $f_\lambda$ is then induced by declaring that 
\be P(\widehat f)~=~\e_{\widehat f} \sigma(\widehat f)\ ,\ee 
where $\sigma$ is an automorphism of the oscillator $\star$-product algebra, that does not act on the component fields and that leaves the sector invariant, and $\e_{\widehat f}=\pm 1$ is referred to as the intrinsic parity of $\widehat f$. Thus, if $\s(\widehat T^\l)=T^{\l'} \sigma_{\l'}^\l$ then 
\be P(f_\l(X,dX))~=~\e_{\widehat f} \s_\l^{\l'} P_{\l'}^{\l''} f_{\l''}(X,dX)\ .\ee
In the models at hand, parity is broken except in the \cite{Sezgin:2003pt}\footnote{Starting from the more general Vasiliev system in which $\widehat D\widehat \Phi=0$ and $\widehat F+i \left(\widehat{\cal V}(\widehat \Phi\star\widehat \kappa,\widehat S_{\una}) \star\widehat J+{\rm h.c.}\right)=0$, where $\widehat{\cal V}$ is a complex deformation function constructed using $\star$-products and trace operations such that the system is integrable \cite{Vasiliev:1990en,Sezgin:2011hq}, and working perturbatively in $\widehat\Phi$, one can show that parity invariance implies either the Type A or the Type B model \cite{Vasiliev:1992av,Sezgin:2003pt,Sezgin:2011hq}.}
\be \mbox{Type A model (parity-even physical scalar)}~:~~P(\widehat A,\widehat \Phi)~=~(\widehat A,\widehat \Phi)\ ,\quad b~=~1\ ,\ee 
\be \mbox{Type B model (parity-odd physical scalar)}~:~~P(\widehat A,\widehat \Phi)~=~(\widehat A,-\widehat \Phi)\ ,\quad b~=~i\ .\ee 

\subsection{Spacetime Formulation}

\paragraph{Projection to ${\cal X}$-Space}

In order project Vasiliev's equations onto a four-dimensional space-time submanifold ${\cal M}\subset {\cal X}$ one assumes 
\be [P_M,\widehat\Phi]_\star~=~0~=~[P_M,\widehat A]_\star\ ,\quad \widehat U^M~=~0\ ,\ee
and realizes the $\star$-product in the basis in which $Y^{\una}\pm Z^{\una}$ become normal-ordered, \emph{viz.}
\be
\label{star product}
\widehat f(Y , Z)  \ \star  \ \widehat g(Y , Z) ~  = ~ \int \frac{d^2\xi d^2\bar  \xi d^2\eta  d^2\bar \eta   }{(2 \pi )^4} e^{ i \eta \xi + i \bar \eta \bar \xi}  \ \widehat f (y+ \xi , \yb + \bar \xi, z + \xi , \zb -  \bar \xi )  \ \widehat g (y+ \eta , \yb+ \bar \eta , z  -  \eta, \zb+ \bar \eta ) \ ,
\ee
where $(\xi, \bar \xi, \eta, \bar \eta)$ are real doublets\footnote{In particular one has $(\widehat f \star \widehat g)^{\dagger}   = \widehat g^{\dagger} \star \widehat f^{\dagger}$ and can use the Principal Value prescription in case of singular integrands.}
which is manifestly Lorentz covariant and in which the symbol of $\widehat\kappa$ is real-analytic at $Y=Z=0$, \emph{viz.}
\be \widehat\kappa~=~\exp(iy^\a z_\a)\ ,\ee
using a simplified notation that does not distinguish between the symbol $\widehat f_{\widehat{\cal O}}$ and $\widehat{\cal O}$. 
One then assumes that all master fields are real analytic in $Y^{\una}$ and $Z^{\una}$ for generic $X^M$, and imposes the initial data 
\be \widehat \Phi|_{Z=0}~=:~\Phi\ ,\quad \widehat U|_{Z=0}~=:~U\ ,\quad \widehat W|_{Z=0}~=:~W\ .\ee
It is furthermore convenient to impose the following Lorentz covariant gauge condition on the twistor-space connection:
\be i_Z \widehat V~=~0\ ,\quad Z~:=~Z^{\una}\partial_{\una}\ . \label{twistorgauge}\ee
One may seek a perturbative expansion of the form
\be \widehat\Phi  ~=~  \sum_{n = 1}^{\infty}    \widehat \Phi^{\{n\}}  \ ,\quad \widehat V  ~=~   \sum_{n=1}^\infty \widehat{V}^{ \{n\}}\ ,\quad \widehat U  ~=~  \sum_{n = 0}^{\infty}  \widehat U^{\{n\}}  \ ,\ee
in which hence
\be \Phi~=~\sum_{n=1}^\infty \Phi^{\{n\}}\ ,\quad U~=~\sum_{n=0}^\infty U^{\{n\}}\ ,\ .\label{pertexp}\ee
Under these assumptions, the equations of motion in ${\cal Z}$-space, \emph{i.e.} in the cokernel of $i_Z$ where $Z:=Z^{\una} \partial_{\una}$, admit the solution
\bea 
\label{Zsolution Phi}
 \widehat \Phi^{ \{n\} }&=& \Phi^{\{n\}}  -  \rho \sum_{n_1+n_2=n } \left[\widehat V^{ \{n_1\} } ,\widehat \Phi^{ \{n_2 \} } \right]_\pi \ ,\\ [5pt ] 
 \label{Zsolution A}
 \widehat V^{ \{n\} } &=& \rho\left(   \frac{i b}{4}  \ d^2z \  \widehat\Phi^{\{n\} } \star \widehat \kappa   +  \frac{ i \bar b }{4}   \ d^2\zb  \  \widehat\Phi^{ \{n\} } \star \widehat{ \bar \kappa}-\sum_{ n_1+n_2=n } \widehat V^{ \{n_1 \} } \star \widehat V^{  \{n_2\} }  \right)\ ,\eea
and
\be \widehat U~=~  \widehat M( U)   \ ,\quad \widehat M~:=~ \left(1+\sum_{n=1}^\infty \widehat L^{\{n\}}\right)^{-1}\ ,\quad L^{\{n\}}(\widehat f)~:=~\rho \left(\left[\widehat V^{\{n\}},\widehat f\right]_\star\right)\ , 
 \ee
which is uniquely defined up to residual gauge transformations, to be specified below, and where the homotopy contracting operator\footnote{For an alternative realization of $\rho$ based on closed contour integrals.}
\bea
\rho   ~:= ~ i_{Z} \frac{1}{{\cal L}_{Z}}  = i_{Z} \int_{ 0 }^{1}\frac{dt}{t} t^{{\cal L}_{Z}} \ .\eea 
Taking into account Bianchi identities, the remaining equations of motion in ${\cal X}$-space, \emph{i.e.} in the kernel of $i_Z$, are equivalent to $(d\widehat U+\widehat U\star\widehat U)|_{Z=0}=0=(d\widehat \Phi+[\widehat U,\widehat \Phi]_\pi)|_{Z=0}$, that is
\be dU^{\{n\}}+J^{\{n\}} ~=~0\ ,\qquad d\Phi^{\{n\}}+P^{\{n\}}~=~0\ ,\label{FDAXspace}\ee
where the sources $J$ and $P$ admit an expansion in terms of multi-linear, totally symmetric functionals in $\Phi$ as follows\footnote{For example, one has 
$J^{\{n\}}=\sum_{k\geqslant 1}
\sum_{m_1+m_2+n_1+\cdots+n_k=n} J^{(k)}(U^{\{m_1\}} U^{\{m_2\}};\Phi^{\{n_1\}}\cdots \Phi^{\{n_k\}})$.
}:
\be J~:=~\sum_{k\geqslant0}J^{(k)}(U^2;\Phi^k)\ ,\quad P~:=~
\sum_{k\geqslant1}P^{(k)}(U;\Phi^k)\ ,\ee
with
\be J^{(k)}~:=~\sum_{k_1+k_2=k} \left.\left(\widehat M^{(k_1)} (U)\star \widehat M^{(k_2)} (U)\right)\right|_{Z=0}\ ,\quad P^{(k)}~:=~\sum_{k_1+k_2=k} \left.\left[\widehat M^{(k_1)} (U), \widehat \Phi^{(k_2)}\right]_\pi\right|_{Z=0}\ ,\ee
where 
\be
\widehat M~:=~ \left(1+\sum_{k=1}^\infty \widehat L^{(k)}\right)^{-1}\ ,\quad L^{(k)}(\widehat f)~:=~\rho \left(\left[\widehat V^{(k)},\widehat f\right]_\star\right)\ , 
 \ee
and $\widehat \Phi^{(k)}$ and $\widehat V^{(k)}$ are give by the recursive relations
\bea 
 \widehat \Phi^{ (k) }&=& \delta_{k1} \Phi  -  \rho \sum_{k_1+k_2=k } \left[\widehat V^{ (k_1) } ,\widehat \Phi^{ (k_2) } \right]_\pi \ ,\\ [5pt ] 
 \widehat V^{ (k) } &=& \rho\left(   \frac{i b}{4}  \ d^2z \  \widehat\Phi^{(k) } \star \widehat \kappa   +  \frac{ i \bar b }{4}   \ d^2\zb  \  \widehat\Phi^{ (k) } \star \widehat{ \bar \kappa}-\sum_{ k_1+k_2=k } \widehat V^{(k_1) } \star \widehat V^{  (k_2) }  \right)\ .\eea
By their construction, these functionals obey the integrability conditions
\be \sum_{k_1+k_2=k}\left(
2J^{(k_1)}(J^{(k_2)}(U^2;\Phi^{k_2})U;\Phi^{k_1})+k_1 J^{(k_1)}(U^2;P^{(k_2+1)}(U;\Phi^{k_2+1})\Phi^{k_1-1})\right)~=~0\ ,\ee
\be \sum_{k_1+k_2=k}\left(P^{(k_1)}(J^{(k_1)}(U^2;\Phi^{k_2});\Phi^{k_1})-k_1 P^{(k_1)}(U;P^{(k_2+1)}(U;\Phi^{k_2+1})\Phi^{k_1-1})\right)~=~0\ ,\ee
which ensure that $U$ and $\Phi$ subject to the generalized curvature constraints \eq{FDAXspace} generate a free differential algebra in ${\cal X}$-space. Defining
\be W~:=~ \widehat W|_{Z=0}\ ,\quad K~:=~\widehat K|_{Z=0}\ ,\ee
the manifest Lorentz covariance implies\footnote{Applying a Lorentz transformation $\delta_{\L}$ with parameter $\L^{\a\b}$ 
to $0\equiv \widehat T+(\omega^{\a\b} \widehat T_{\a\b}-{\rm h.c.})$ where $\widehat T:=
\widehat W-\widehat M(W)$ and $\omega^{\a\b} \widehat T_{\a\b}-{\rm h.c.}:=\widehat M
(K)-\widehat K$, it follows that $\delta_{\L} \widehat T+\left((d\L^{\a\b}+2\L^{\a\c}\o_{\c}
{}^\b) \widehat T_{\a\b}+\omega^{\a\b}\delta_{\L}\widehat T_{\a\b}-{\rm h.c.}\right)\equiv 
0$ for all $\L^{\a\b}$, that is, $\delta_{\L}\widehat T\equiv 0$, $\delta_{\L}\widehat T_{\a
\b}=2\L_{(\a}{}^\c \widehat T_{\b)\c}$ and $\widehat T_{\a\b}\equiv 0$. Thus $\widehat M
(K)-\widehat K\equiv 0$, as one can verify order by order in $\Phi$-expansion treating $
\o^{\a\b}$ as a zeroth order object; for example, $(\widehat M(K)-\widehat K)^{(1)}=K^
{(1)}-\widehat L^{(1)}(K^{(0)})-\widehat K^{(1)}$ where $\widehat L^{(1)}(K^{(0)})=\int_0^1 
dt z^\a\left[\widehat V^{(1)}_\a,K^{(0)}\right]_\star-{\rm h.c.}$ and $\widehat K^{(1)}=-\frac12 \o^{\a
\b} \{z_\a,\widehat V^{(1)}_\b\}_\star-{\rm h.c.}$ with $K^{(0)}=\frac1{4i} (\o^{\a\b}y_{\a} y_
\b+{\rm h.c.})$ and $\widehat V^{(1)}_\a=-\frac{ib}2 z_\a \int_0^1 t\,dt \,\Phi(-tz,
\yb)e^{ityz}$, which can be rewritten as $\widehat L^{(1)}(K^{(0)})=-\widehat K^{(1)}\propto \o^{\a\b} z_\a z_\b \int_0^1 t(1-t) \, dt \,\Phi(-tz,\yb)e^{ityz}$. Thus, also $K^{(1)}=0$ and hence $
(\widehat M(K)-\widehat K)^{(1)}\equiv 0$.} 
$\widehat W \equiv \widehat M(W+K)-\widehat K\equiv \widehat M(W)$. Thus, the manifestly Lorentz covariant form of the generalized curvature constraints in ${\cal X}$-space reads
\be \nabla W+W\star W+r+\Big(\!\!\!\!\!\!\!\!\!\sum_{\tiny\ba{c} k\geqslant 1\\ k_1+k_2=k\ea} \!\!\!\!\!\!\!\widehat M^{(k_1)} (W)\star \widehat M^{(k_2)} (W)-i\!\!\!\!\!\!\!\!\!\sum_{\tiny\ba{c} k\geqslant 2\\ k_1+k_2=k\ea} \!\!\!\!\!\!\!(r^{\a\b}\widehat V^{(k_1)}_\a\star \widehat V^{(k_2)}_\b+{\rm h.c.})\Big)\Big|_{Z=0}~=~0\ee
\be \nabla \Phi+\left[W,\Phi\right]_\pi+\!\!\!\!\!\!\!\!\!\sum_{\tiny\ba{c} k\geqslant 2\\ k_1+k_2=k\ea} \!\!\!\!\!\!\! \left[\widehat M^{(k_1)} (U), \widehat \Phi^{(k_2)}\right]_\pi\Big|_{Z=0}~=~0\ .\ee

\paragraph{Residual Gauge Symmetries}

The generalized curvature constraints on ${\cal X}$ remain invariant under residual gauge transformations $\delta_{\widehat\e_{\rm res}}$ with parameters obeying
\bea
\label{e twistor condition}
{\cal L}_{Z}\widehat \e_{\rm res}   + i_Z \left[ \widehat V , \widehat \e_{\rm res}  \right]_{\star }  ~=~  0  \ ,
\eea
as to preserve the twistor space gauge condition $i_Z\widehat V=0$. Under the assumption of real analyticity in twistor space and the existence of a $\Phi$-expansion, one finds
\bea
\label{e perturbative}
\widehat \e_{\rm res}  ~=~ \widehat M(\e)\ ,  \quad \e ~ :=~ \widehat \e_{\rm res}|_{Z=0}\ .\eea
The action of $\delta_{\widehat\e_{\rm res}}$ on $\widehat\Phi$ and $\widehat U$ is equivalent to acting directly on the initial data $\Phi$ and $U$ as follows:
\be 
\delta_{\e}U~=~\sum_{k\geqslant 0} \delta^{(k)}_\e U\ ,\quad \delta^{(k)}U~:=~\delta_{k0}d\e+\sum_{k_1+k_2=k} \left[\widehat U^{(k)},\widehat \e_{\rm res}\right]_\star\Big|_{Z=0}\ ,\ee
\be \delta_{\e}\Phi~=~\sum_{k\geqslant 1} \delta^{(k)}_\e \Phi\ ,\quad \delta^{(k)}\Phi~:=~-\sum_{k_1+k_2=k} \left[\widehat\e^{(k)}_{\rm res},\widehat \Phi^{(k_2)}\right]_\pi\Big|_{Z=0}\ .\ee

\paragraph{Formulation in Four-Dimensional Spacetime}

Using the shift symmetry to set $\frac{\partial^2}{\partial y^\a \partial y^\b}W|_{Y=0}=0$, 
one may write 
\be W=e + w\ ,\quad w~:=~\sum_{s=4,6,\dots} w_{s}\ ,\ee
with 
\be e~=~\frac\l{2i} e^{\a\ad} y_\a \yb_
{\ad}\ ,\quad w_{s}~=~\frac1{2i} \sum_{t=0}^{s-1}w_{\{s-1,t\}}\ ,\ee
where 
\bea w_{\{s-1\}}&=& w^{\a
(s-1)\ad(s-1)} y^{s-1}_{\a(s-1)} \yb^{s-1}_{\ad(s-1)}\ ,\\[5pt]w_{\{s-1,t\}}&=&w^{\a
(s-1+t)\ad(s-1-t)} y^{s-1+t}_{\a(s-1+t)} \yb^{s-1-t}_{\ad(s-1-t)}+{\rm h.c.}\ ,\eea
for $t\geqslant 1$, in a notation in which $\{s_1,s_2\}$ denote highest weights of (real) 
tensors of the Lorentz $so(3,1)$. Likewise, one can expand 
\be \Phi~=~\sum_{s=0,2,4,\dots} 
\Phi_s\ ,\quad \Phi_s~=~\sum_{n=0}^\infty \Phi_{\{s+n,s\}}\ ,\ee
where 
\bea \Phi_{\{n\}}&=&\Phi^{\a(n)
\ad(n)}y^{n}_{\a(n)} \yb^n_{\ad(n)}\ ,\\[5pt]\Phi_{\{s+n,s\}}&=&\Phi^{\a(2s+n)\ad(n)}y^{s+n}_{\a
(s+n)} \yb^n_{\ad(n)}+\pi({\rm h.c.})\ ,\eea
for $s\geqslant 2$. This yields 
\bea \nabla e+ r+e\star e +(\{\widehat M^{(1)}(e),e\}_\star)_2 + T_2&=&0\ ,\\[5pt]
 \nabla w_s+ \{e,w_s\} +(\{\widehat M^{(1)}(e),e\}_\star)_s + T_s&=&0\ ,\\[5pt]
\nabla \Phi_s+[e,\Phi_s]_\pi+S_s&=&0\ ,\eea
where $T_s$ and $S_s$ denote the projections to the spin-$s$ sectors of adjoint and twisted-adjoint sources $T$ and $S$, respectively, given by
\be T~=~\sum_{\ell\geqslant 2} T^{\langle \ell\rangle }\ ,\qquad 
S~=~\sum_{\ell\geqslant 2} S^{\langle \ell\rangle }\ ,\ee
with
\be T^{\langle \ell\rangle }:=\!\!\sum_{\ell_1+\ell_2=\ell}\!\!\Big(\!\!\left(\widehat M^{(\ell_1)}(e)+\widehat M^{( \ell_1-1)}(w)\right)\star \left(\widehat M^{(\ell_2)}(e)+\widehat M^{( \ell_2-1)}(w)\right) -i (r^{\a\b}\widehat V^{(\ell_1)}_\a\star \widehat V^{(\ell_2)}_\b+{\rm h.c.})\Big)\Big|_{Z=0}\ ,\ee
\be S^{\langle \ell\rangle }~:=~\sum_{\ell_1+\ell_2=\ell}\left[\widehat M^{(\ell_1)}(e)+\widehat M^{( \ell_1-1)}(w),\widehat \Phi^{(\ell_2)}\right]_\pi\Big|_{Z=0}\ .\ee
On four-dimensional submanifolds where $e$ is invertible,  
the field content can be expressed in terms of the dynamical fields
\be \phi~:=~\Phi_{\{0,0\}}\ ,\quad g_{\mu\nu}~:=~ e_\mu^{\a\ad} e_{\nu,\a\ad}\ ,\quad \phi_{\mu(s)}~:=~(e_{\mu}^{\a\ad})^{s-1} w_{\mu,\a(s-1)\ad(s-1)}\ ,\ee
with $s=4,6,\dots$ in the minimal bosonic models and $s=1,3,4,\dots$ in the non-minimal bosonic models, and gauge functions corresponding to shift symmetries, by treating $g_{\mu\nu}:=e^a_\mu e^b_\nu \eta_{ab}$ and $(\omega_\mu^{\a\b},\bar \omega_\mu^{\ad\bd})$ as zeroth order fields and $\phi_{\mu(s)}$ as weak fields together with all components in $
\Phi$, whereby $T^{\langle \ell\rangle}$ and $S^{\langle \ell\rangle}$ have weak-field 
expansions starting at the $\ell$th order. 
More precisely, all non-dynamical fields can be expressed in terms of the dynamical fields and their derivatives except the $\{s-1,t+1\}$ component of $w_{s-1,t}$ for all $s$, on which the residual gauge symmetries with parameters $\e_{\{s-1,t+1\}}$ act as shift symmetries. 
In the parity invariant models, the dynamical scalar field has parity $P(\phi)=b^2 \phi$, \emph{i.e.} it is a scalar and a pseudo-scalar in the Type A and B models, respectively, while all fields with spin $s\geqslant 1$ have intrinsic parities $+1$, \emph{i.e.} they transform as tensors (and not pseudo-tensors) under $P$\footnote{In the harmonic expansion, the states in the lowest-weight space $D(s+1;(s))$ have parity $(-1)^s$.}.

\paragraph{Strongly Coupled Derivative Expansions and Scheme Dependencies}

The resulting dynamical field equations in four dimensions are manifestly generally covariant and have double expansions in weak-
fields and derivatives.
In the first order in weak fields, one has kinetic terms that are second order in 
derivatives and that contain critical mass terms defining an inverse anti-de Sitter radius, $\l$ say.
In the higher orders, one has generalized stress tensors that contain all possible orders in derivatives weighted with $\l^{-1}$.
In other words, treated in a perturbative expansion around the anti-de Sitter vacuum, the spacetime formulation contains strongly coupled derivative expansions.

As for the higher spin gauge symmetries with residual parameters $\e_{\{s-1\}}$, $s\geqslant 1$, the corresponding kinetic terms obey generalized Bianchi identities and hence the stress tensor obey generalized on-shell conservation laws\footnote{For $s=1,2$, the kinetic terms obey standard Bianchi identities while for $s\geqslant 3$ their divergencies are equal to constructs whose weak-field expansions start in the second order; thus the spin-$1$ current and the spin-$2$ stress tensor are conserved while the divergencies of spin-$s$ stress tensors for $s\geqslant 3$ contain a classical anomaly that cancel the divergence of the kinetic term.}. 
In order to examine these, a careful analysis is required, however, since in taking the divergence of the spin-$s$ equation of motion ($s\geqslant 1$), and rearranging derivatives, each given tensorial structure, with fixed numbers of fields, derivatives and $\l$, receives infinitely many contributions, that are all of the same order; thus it is conceivable that higher spin gauge invariance can be verified only formally at the level of suitably analytically continued ``coupling functions''.

Thus, it remains an open problem whether there exists a globally defined formulation of higher spin gravity directly in terms of component fields in four-dimensional spacetime\footnote{Possibly, such a formulation may be reachable by means of a field redefinition as to remove the strongly coupled derivative expansions; for a recent review, see \cite{Bekaert:2010hw}. Indeed, at the cubic level, the most general vertices in anti-de Sitter spacetime that involve three massless fields with spins $s_1$, $s_2$ and $s_3$ and that are cohomologically nontrivial in the Fronsdal program, involve only up to $s_1+s_2+s_3$ derivatives, which appear in the three-curvature vertex. The Fronsdal program is, howver, far from completed. In fact, as pointed out already in \cite{Vasiliev:1990en,Vasiliev:1992av}, it is conceivable that the constructive approach is to exploit the somewhat arbitrary choices that have been made in imposing \eqref{twistorgauge} and embedding ${\cal X}$ into ${\cal C}$ at $Y^{\una}=Z^{\una}=0$.}.
In view of these subtleties, to which one may also add the dependence on the choice of the twistor gauge \eqref{twistorgauge} as well as the specific realization \eqref{star product} of the $\star$-product, which is by no means the unique manifestly Lorentz covariant choice for which the symbol of $\widehat\kappa$ is real-analytic, it makes sense to examine to what extent physical observables that do not rely crucially on nontrivial spacetime topology, such as amplitudes in perturbative expansions around anti-de Sitter spacetime, can be extracted using a suitable dual twistor-space method that is manifestly invariant under both higher spin gauge transformations and re-definitions of symbols.

\subsection{Twistor Space Formulation}

\paragraph{Gauge Functions}

The flatness condition on $\widehat U$ implies the existence of locally defined gauge functions
\be \widehat L~:=~\widehat L'\star \widehat L_0\ ,\quad \widehat{L}_0 ~:=~ {\rm P}\left[ \exp_\star \int_{p_0}^p\widehat U\right]\ ,\quad p,p_0~\in~{\cal X}_I\ ,\ee
where ${\rm P}\left[ \cdot\right]$ denotes path ordering, such that
\be
\label{space-time dependence1}
 \widehat U ~=~ \widehat L^{-1} \star  d \widehat L\ , \quad 
  \widehat \Phi   ~=~ \widehat L^{ - 1}   \star  \widehat \Phi '    \star    \pi(  \widehat L) 
\ ,\quad \widehat V  ~=~  \widehat L^{ - 1}   \star  ( \widehat V'  +  q ) \star\ \widehat L  \ , 
\ee
where the primed master fields obey $d\widehat \Phi '   = 0=d\widehat V'$ and 
\be q\widehat V'+\widehat V'\star\widehat V'~=~\frac{i}4 \left(b\, dz^2  \widehat \Phi'\star\widehat \kappa+\bar b \,d\zb^2 \widehat \Phi'\star\widehat{\bar\kappa}\right)\ ,\quad q\widehat \Phi' +[\widehat V',\widehat\Phi']_\pi~=~0\ .\label{twistorspaceequations}\ee
Assuming real analyticity of the primed fields\footnote{
This assumption brings a loss of generality, as it may be convenient, for example in order to manifestly exhibit symmetries of solutions, to choose a base point $p_0$ at which the full master fields are singular; for example, in the case of spherically symmetric solutions \cite{Iazeolla:2011cb} it is convenient to place the base point at the origin, where $\widehat\Phi'$ is given by a sum of delta functions on ${\cal Y}$ and their derivatives, and $\widehat V'$ has an algebraic singularity along a plane in ${\cal Y}\times {\cal Z}$ (as these singularities intersect the point $Y=0=Z$, they show up as spacetime singularities in the component fields at the origin).}, 
the gauge condition \eqref{twistorgauge} is equivalent to that
\be \widehat L~=~L-\rho(\Upsilon_{\widehat L,\widehat V''})\ ,\label{widehatLeq}\ee
where
\be L~:=~\widehat L\Big|_{Z=0}\ ,\quad \widehat\Upsilon_{\widehat L,\widehat V'}~:=~\left((\widehat L^{-1}-1)\star q\widehat L+\widehat L^{-1}\star \widehat V'\star \widehat L\right)\ ,\ee
as can be seen from $0=\rho (\widehat V)=\rho(\widehat L^{-1}\star (\widehat V'+q)\star \widehat L)=\rho (q \widehat L)+\rho(\widehat \Upsilon_{\widehat L,\widehat V'})$ where $\rho (q \widehat L)\equiv\widehat L-L$. Assuming that $\widehat L'$ can be chosen such that\footnote{The assumption is equivalent to that $\widehat V^{\prime\prime}:=(\widehat L')^{-1}\star(\widehat V'+q)\star \widehat L'=\sum_{k\geqslant 0}\widehat V^{\prime\prime(k)'}$ and $\widehat \Phi^{\prime\prime}:=(\widehat L')^{-1}\star \widehat \Phi'\star \pi(\widehat L')=\sum_{k\geqslant 1}\widehat \Phi^{\prime\prime(k)'}$ where $\widehat V^{\prime\prime(0)'}:=(\widehat L')^{-1}\star q\widehat L'$. Conversely, starting from $\widehat V^{\prime\prime}:=\sum_{k\geqslant 0}\widehat V^{\prime\prime(k)'}$ and $\widehat \Phi^{\prime\prime}:=\sum_{k\geqslant 1}\widehat \Phi^{\prime\prime(k)'}$ where thus $\widehat V^{\prime\prime(0)'}$ obeys $q\widehat V^{\prime\prime(0)'}+\widehat V^{\prime\prime(0)'}\star \widehat V^{\prime\prime(0)'}=0$, it may be the case $\widehat V^{\prime\prime(0)'}$ cannot be expressed using a gauge function. In this special case, the gauge $i_Z \widehat V=0$ cannot be imposed using the recursive method explained here. 
}
\be \widehat \Phi'~=~\Phi'+\sum_{k\geqslant 2} \widehat \Phi^{\prime(k)'}\ ,\quad  \widehat V'~=~\sum_{k\geqslant 1} \widehat V^{\prime(k)'}\ ,\ee
where $\widehat \Phi^{\prime(k)'}$ and $\widehat V^{\prime(k)'}$ are recursively defined $k$-linear symmetric functionals in $\Phi'$, it follows that
\be \widehat L~=~L+\sum_{k\geqslant 1}\widehat L^{(k)'}\ .\ee

\paragraph{Different Types of Moduli}

Thus, in order to solve the Vasiliev equations one may use the following gauge function method:
\begin{itemize}
\item \emph{Step (i)}: Compute\footnote{For example, one may apply recursive homotopy contraction some exact method such as separation of variables.} $\widehat V'$ and $\widehat \Phi'$ from the twistor-space equations \eqref{twistorspaceequations} with $\Phi'$ in a suitable sector of the theory;
\item \emph{Step (ii)}: Compute $\widehat L$ from \eqref{widehatLeq};
\item \emph{Step (iii)}: Choose $\widehat L'$ depending on boundary conditions in ${\cal Z}$;
\item \emph{Step (iv)}: Choose $L$ depending on boundary conditions in ${\cal X}$;
\end{itemize}
which thus introduces three types of moduli, namely $\Phi'(Y)$ and the boundary values of $\widehat L'(Z,Y)$ and $L(X,Y)$.

\subsection{Zero-Form Charges}

In what follows, we review the basic properties of zero-form charges and their  interpretation as basic building blocks for on-shell amplitudes, that we refer to as quasi-amplitudes.

\paragraph{Globally Defined Formulations and Observables}

Given a base manifold ${\cal M} = \bigcup_{I} {\cal M}_{I}$, where ${\cal M}_I$ are simply connected coordinate charts, and locally defined unfolded systems consisting of zero-forms $\Phi^i_I$ and one-forms $W^r_I$ obeying Cartan integrable equations of motion
\be
d\Phi^i+W^r Q^i_r(\Phi)~\approx~0 \ ,\quad dW^r+W^s W^t f^r_{st}(\Phi)~\approx~0\ ,
\ee
a globally defined formulation arises by gluing together the locally defined configurations across the chart boundaries using transition functions 
\be T_I^J:(\Phi^i_J,W^r_J)|_{{\cal M}_I\cap {\cal M}_J}\rightarrow (\Phi^i_I,W^r_I)|_{{\cal M}_I\cap {\cal M}_J}\ ,\ee
valued in a given subgroup of the group of Cartan gauge transformations, referred to as the structure group; for further details, see for example \cite{Boulanger:2011dd, Sezgin:2011hq,Boulanger:2012bj} and references therein. 
Thus, the one-forms decompose into $W^r=(\Gamma^\a;E^a)$ where $\Gamma^\a$ are connections with inhomogeneous transitions, and $E^a$ are soldering forms that  together with $\Phi^i$ form sections associated to the principal bundle of the structure group. 

A given unfolded system may thus admit many inequivalent globally defined formulations, which one may think of as different phases of the theory, referred to as homotopy phases in \cite{Colombo:2010fu}, each characterized by its structure group, or equivalently,
by the set of classical observables that break the Cartan gauge group down to the structure group off shell.
The latter, which one may think of as order parameters,
are thus functionals ${\cal O}[\Phi,E]$ obeying $\delta_{\e_\L}{\cal O}=0$ off shell for unbroken gauge parameters $\e^r_\L=(\L^\a;0)$, which can be defined locally, and $\delta_{\e_\xi}{\cal O}\approx 0$ for gauge parameters $\e^r_\xi=(0;\xi^a)$ forming sections, which one may refer to as topologically broken, or softly broken, gauge parameters.

Using the Cartan integration technique \cite{Colombo:2010fu,Boulanger:2012bj}, the locally defined configurations can be expressed on shell as finite gauge transformations of integration constants $\{\Phi^{\prime i}_I\}$ generated by gauge functions $\{\l^r_I\}=\{(0;\l^a_I)\}$, where one may identify $\Phi^{\prime i}_I=\Phi^i_I|_{p_I}$ at base points $p_I\in {\cal M}_I$ where $\l^a_I|_{p_I}=0$. The invariance properties of ${\cal O}$ implies that the gauge functions can be switched off in the interior of ${\cal M}$ and that the charts can be deformed such that the base points can be moved together to a single base point $p_0$, such that
\be {\cal O}[\Phi,E]~\stackrel{\rm Cartan~ int.}{\approx}~{\cal O}[\Phi(\Phi',\l),E(\Phi',\l,d\l)]~\stackrel{\rm deform}{=:}~{\cal S}[\Phi|_{p_0},\l|_{\partial {\cal M}}]\ ,\label{quasi-action}\ee
which one may think of as a contribution to an on-shell action, that we shall refer to as a quasi-action. 
In higher spin gravity, the quantities $\Phi|_{p_0}$ and $\l|_{\partial {\cal M}}$, respectively, are represented by $\Phi'(Y)$ and the boundary values of $L(X,Y)$, as discussed above, and the construction of quasi-actions has been initiated in \cite{Colombo:2010fu,Sezgin:2011hq}, as we shall discuss below, after we have first introduced the notion of zero-form charges.

\paragraph{Zero-Form Charges and Quasi-Amplitudes}

We refer to classical observables that can be evaluated using the field content of a single coordinate chart as being locally accessible \cite{Colombo:2010fu}. 
A particular class of such observables are composite zero-forms ${\cal I}(\Phi)$ that are closed on shell, that is
\be d {\cal I}~\approx~ - W^rQ^i_r \partial_i {\cal I}~=~0\ ,\ee
or equivalently, that are invariant under general Cartan gauge transformations, that is
\be \delta_\e {\cal I}~=~-\e^r Q^i_r \partial_i {\cal I}~=~0\ ,\ee
as both properties are equivalent to $Q^i_r\partial_i {\cal I}=0$. 
As ${\cal I}$ is globally defined for any choice of structure group, it is observable in any phase of the theory, and as it is closed on shell, its value on a given solution, which we refer to as the zero-form charge, can be evaluated at any point in ${\cal M}$.
These charges thus provide locally accessible quasi-actions that are independent of gauge functions\footnote{The existence of zero-form charges is closely related to the fact that in unfolded dynamics an infinite tower of zero-forms is required in order to describe each local degree of freedom. Realizing these towers as functions on fiber spaces, the zero-form charges become integral over these fibers at a given point on the base manifold, which one may thus think of as a contribution to a dual fiber action.}. 

Assuming a $\Phi$-expansion
\be Q^i_r~=~\sum_{n\geqslant 0} Q^{(n)i}_r\ ,\quad f^r_{st}~=~\sum_{n\geqslant 0} f^{(n)r}_{st}\ ,\quad {\cal I}~=~\sum_{n\geqslant 0}{\cal I}^{(n+n_0)}\ ,\ee
where ${\cal I}^{(n)}$ are $n$-linear functionals of $\Phi$ obeying  equivariance relations
\be \sum_{n_1+n_2=n} \delta^{(n_1)}_{\e} \partial_i {\cal I}^{(n_2)}~=~0\ ,\quad \delta^{(n)}\Phi^i~:=~\e^r Q^{(n)i}_r\ ,\ee
it follows that the leading contribution ${\cal I}^{(n_0)}$ is invariant under the undeformed Cartan gauge algebra, \emph{viz.}
\be \delta^{(0)}_{\e} {\cal I}^{(n_0)}~=~0\ ,\quad [\delta^{(0)}_{\e_1},\delta^{(0)}_{\e_1}]~=~\delta^{(0)}_{\e_{12}}\ ,\quad \e^r_{12}~=~f^{(0)r}_{st}\e^s_1 \e^t_2\ .\ee
Assuming that the undeformed gauge algebra contains the spacetime isometry algebra, as is the case in unfolded theories of gravity, and given a set of connected zero-form invariants $\{{\cal I}_{n_0}\}$, one may thus ask for sectors of states\footnote{Sectors of states labelled by points can be constructed by acting with undeformed gauge group elements $L(p,s;\ell)$ on a reference state $\Phi'_0$, centered at $p_0$, that is, $\Phi'_i=\left.\left(\rho^{(0)}(L_i)\Phi_0\right)\right|_{p_0}$.
} $\Phi'_i\equiv \Phi'(p_i,s_i;\ell_i)$, labelled by points $p_i$, Lorentz spins $s_i$ and other internal labels $\ell_i$, such that all ${\cal I}_{n_0}^{(n_0)}(\Phi'_1,\dots,\Phi'_{n_0})$ fall off sufficiently fast in limits where all $p_i$ are separated far enough from each other, in which case one may refer to these quantities as quasi-amplitudes \cite{Colombo:2010fu}.

\paragraph{Zero-Form Charges in Higher Spin Gravity}

In higher spin gravity, a natural set of intrinsically defined and manifestly gauge invariant observables are given by decorated Wilson loops \cite{Sezgin:2011hq}
\be {\cal W}(\gamma;\{p_i\})~=~\widehat {\rm Tr}\left[P\left[e_\star^{\oint_\gamma \widehat U}\prod_i \widehat {\cal V}_i(\widehat \Phi,\widehat S_{\una})|_{p_i}\right]\right]\ ,\ee
along closed paths $\gamma\subset {{\cal M}}$ with adjoint impurities $\widehat{\cal V}_i$ inserted at points $p_i\in \gamma$; the flatness of $\widehat U$ and covariant constancy of adjoint $\star$-functions of $(\widehat\Phi,\widehat S_{\una})$ ensures that ${\cal W}$ is invariant under smooth deformations of $\gamma$ and $p_i$. 
The basic impurity is given by the vertex operator
\be 
\widehat{\cal V}_{k,\bar k}(\m,\bar \m)~=~e_\star^{i( \m\widehat S-\bar\m\widehat {\bar S)}}\star
(\widehat \Psi)^{\star k}\star (\widehat{\bar\Psi})^{\star \bar k}\ ,\label{vertex}
\ee
where $(\m^\a,\bar\m^{\ad})$ are auxiliary twistor momenta, $k,\bar k\in\mathbb N$ and we have introduced 
\be 
\widehat \Psi~:=~\widehat \Phi\star\widehat\kappa\ ,\qquad \widehat{\bar\Psi}~:=~\widehat \Phi\star \widehat{\bar\kappa}\ ,
\ee
obeying the following relations
\be d\widehat \Psi+[\widehat U,\widehat\Psi]_\pi~=~0\ ,\qquad d\widehat {\bar \Psi}+[\widehat U,\widehat {\bar \Psi}]_\star~=~0\ ,\label{dpsi}\ee
\be \{\widehat S_\a,\widehat \Psi\}_\star~=~[\widehat S_\a,\widehat {\bar \Psi}]_\star~=~0\ ,\qquad \{\widehat S_{\ad},\widehat {\bar\Psi}\}_\star~=~[\widehat S_{\ad},\widehat \Psi]_\star~=~0\ ,\label{sss2}\ee
\be \widehat \Psi\star \widehat \Psi ~=~\widehat{\bar \Psi}\star \widehat{\bar \Psi}\ ,\qquad [\widehat\Psi,\widehat{\bar\Psi}]_\star~=~0\ ,\qquad \widehat \Psi\star \widehat\kappa\widehat{\bar\kappa}~=~\widehat{\bar\Psi}\ . \label{psipsi}
\ee
Thus, if $\gamma$ is trivial and if all the impurities can be moved to a single point $p_0$, the decorated Wilson loops collapse to the zero-form charges ($\s=0,1$)
\bea 
{\cal I}^\sigma_{k,\bar k}(\m,\bar\m)~=~\widehat{\rm Tr}\left[(\widehat\kappa\widehat{\bar{\kappa}})^\s\star \widehat{\cal V}_{k,\bar k}(\mu,\mb)\ \right]\ .\eea
In the parity invariant models, one has
\be P({\cal I}^\sigma_{k,\bar k}(\m,\bar\m))~=~b^{2(k+\bar k)}{\cal I}^\sigma_{\bar k, k}(\mb,\m)\ .\ee
The following zero-form charges are algebraically independent:
\bea
\label{charge unprime}{\cal I}_{n_0}(\mu,\bar\mu)&:=&\widehat{\rm Tr}\left[(\widehat\Phi\star\widehat\kappa)^{\star(n_0)}\star \widehat{\cal V}(\mu,\mb)\star \widehat\kappa\widehat{\bar\kappa}\right]\ ,\qquad n_0~=~2,4,\dots\ ,\\[5pt]
\label{charge prime}
{\cal I}'_{n_0}(\mu,\bar \mu)&:=&\widehat{\rm Tr}\left[(\widehat\Phi\star\widehat{\bar\kappa})^{\star(n_0)} \star \widehat{\cal V}(\mu,\mb) \right]\ ,\quad\qquad n_0~=~1,2,3,\dots\ ,\label{106}
\eea
where $\widehat{\cal V}(\mu,\mb):= \widehat{\cal V}_{0,0}(\mu,\mb)$. 
We shall assume that if $\left.\frac{\partial}{\partial \mu^{\a_1}}\cdots \frac{\partial}{\partial \mu^{\a_m}} {\cal I}^\sigma_{k,\bar k}(\m,\bar\m) \right|_{\m=0}$ is finite then it vanishes for $m>0$ \cite{Sezgin:2011hq}, 
as can be shown formally by making use of the deformed oscillator algebra and the cyclicity property of the trace\footnote{The cyclicity of the trace holds formally modulo boundary terms in twistor space that we assume vanish if the trace is finite.}. 

\section{Regularization Using Twistor Space Plane Waves}\label{Sec:reg}

The zero-form charges are given by integrals over twistor space that may in principle be divergent in a given sector of states.
In \cite{Colombo:2010fu}, it was found that in the basis of twistor-space plane waves and using the normal-ordered form \eqref{star product} of the $\star$-product, the perturbative expansion of ${\cal I}_{n_0}(0,0)$ is well-defined, and arguments were presented supporting the claim that ${\cal I}^{(n)}_{n_0}(0,0)$ vanishes if $n>n_0$.
Moreover, as for ${\cal I}'_{n_0}(0,0)$, it was proposed to remove its leading divergence 
\bea
 {\cal I}^{\prime(n_0)'}_{n_0}(0,0)&=&  \widehat{\rm Tr} \left[ (\Phi^{\prime} \star\widehat\kappa)^{\star(n_0)}\right]~\sim~ \left\{ \ba{ll}  \int \frac{d^2z}{2\pi}\ ,&\mbox{$n_0$ odd}   \\[5pt]
\int \frac{d^2z d^2\zb}{(2\pi)^2} \ ,&\mbox{$n_0$ even}\ea\right. \ ,
\eea
by means of a naive multiplicative renormalization, which suppresses, however, all sub-leading corrections.
In what follows, we shall instead propose a refined regularization scheme based on smearing ${\cal I}'_{n_0}(\mu,\mb)$ with suitable regularization functions of $(\mu,\bar\mu)$ 
as to extract well-defined quasi-amplitudes including sub-leading corrections.

\subsection{Regularization Functions}

We shall assume that the lemma below Eq. \eqref{106} applies to the plane-wave sector, 
so that we have
\bea {\cal I}_{n_0}(\mu,\bar \mu)&=&{\cal I}_{n_0}(0,0)\ ,\\[5pt]
\label{ODD1}  {\cal I}'_{n_0}(\mu,\bar \mu) &=& {\cal I}_{n_0}(0, \mb)  \ ,\qquad n_0~=~1,3,\dots\ . \eea
A scheme for regularizing ${\cal I}'_{n_0}(0,0)$ in the plane-wave sector that keeps sub-leading corrections, is to consider the smeared invariants 
\be
\label{reg}
\cI_{n_0}|_{\rm reg} ~:=~\int\frac{d^2\mu d^2\mb}{(2\pi)^2} \widetilde {\cal V}_{n_0}(\mu,\mb) {\cal I}_{n_0}(\mu,\bar \mu) ~\equiv ~\widehat{\rm Tr}\left[(\widehat\Phi^{\prime} \star\widehat \kappa )^{\star(n_0)}\star \widehat{\cal V}_{n_0}(\widehat { S }'_{\una}) \right]\ ,
\ee
where the twistor-space regulators 
\be
\widehat{\cal V}_{n_0}(\widehat S'_{\una})  ~:=~  \int \frac{d^2\m d^2\mb }{(2 \pi)^2 } 
\widehat{\cal V}(\m,\bar \m) \widetilde {\cal V}_{n_0}(\mu,\bar\mu)~=~\int \frac{d^2\m d^2\mb }{(2 \pi)^2 } 
e_\star^{i(\mu\widehat S'-\mb\widehat{\overline S}')}\widetilde {\cal V}_{n_0}(\mu,\bar\mu)\ ,\ee
with 
\be  \label{ODD2}\widetilde {\cal V}_{n_0}(\mu,\bar\mu)~:=~ \delta^2(\m) \widetilde {\cal V}_{n_0}(\mb)\ ,\quad \widehat{\cal V}_{n_0}(\widehat S'_\ad)~:=~ \widehat{\cal V}_{n_0}(0, \widehat S'_\ad)~=~\int \frac{d^2\mb }{2 \pi } 
e_\star^{- i\mb\widehat S'}\widetilde {\cal V}_{n_0}(\mb)\      ,\quad \mbox{for $n_0=1,3,\dots$}\ ,  \ee
to be determined by requiring that the zero-form charges
\begin{itemize}
\item[i)]  are finite and gauge invariant on shell;
\item[ii)] can be treated as meaningful deformations of the generalized Hamiltonian bulk action of \cite{Boulanger:2011dd}.
\end{itemize}
Condition (i) is to be examined in what follows, while a procedure for implementing (ii) is proposed in Section \ref{Sec:action}. 

In particular, in the leading order, condition (i) requires that $\widetilde{\cal V}_{n_0}(0)$ (odd $n_0$ ) and $\widetilde{\cal V}_{n_0}(0,0)$ (even $n_0$) are finite.

\subsection{Twistor Space Plane Waves}

The zero-form charges and the related quasi-amplitudes can be expanded in twistor-space plane waves, \emph{viz.}  
\be
{\cal I}^{(n)'}(\Phi'_1, \dots, \Phi'_n)   =  \left( \prod_{k = 1}^{n} \int  \frac{ d^4 \L_k}{(2 \pi)^2} \widetilde \Phi'_i(\L_i ) \right)  \widetilde{{\cal I}}^{(n)'} (\L_1,\dots, \L_n) \ ,
\ee
where ${\cal I}^{(n)}$ denotes the $n$th order correction to a generic zero-form charge ${\cal I}$ and 
\be \widetilde {{\cal I}}^{(n)'} (\L_1,\dots, \L_n)~:=~{{\cal I}}^{(n)'} (\Phi_{\L_1},\dots, \Phi_{\L_n})\ ,\ee
and we use the chiral Fourier transform \footnote{We have changed the sign convention with respect to \cite{Colombo:2010fu}}
\be
\widetilde \Phi'(\L )~  :=~ \int_{\mathbb  R^2 \times \mathbb  R^2 } \frac{ d^4U }{(2 \pi)^2}  \Phi_{  - \L}(U)  \, \Phi'( U)  \ , \quad \Phi_{\L}(U)~:=~ e^{i( \l u -  \lb \bar u)} \ .     \label{twistor wave}
\ee
The inverse transformation reads
\be
\Phi'( Y ) ~=~  \int \frac{ d^4 \L}{(2 \pi)^2} \  \Phi_{\L}(Y) \  \widetilde \Phi'( \L ) \ , 
\ee
where $\L^{\una}:=(\l^\a,-\bar \l^{\ad})$ and $\l^\a$ and $\bar \l^{\ad}$ are treated as two independent real doublets, and we use the normalisations 
\be
\int \frac{d^2 u}{(2 \pi)^2} e^{i \l u } ~=~ \d^2(\l)  \ , \quad \int \frac{d^2\bar u }{(2 \pi)^2}  e^{- i \lb \bar u }  ~=~  \d^2(\lb) 
\ .
\ee 
 The hermitian conjugation $\dagger$, the auto-morphisms $\pi$ and $\bar\pi$ and the anti-automorphism $\tau$ are extended as follows: 
\be (\l^\a,\bar\l^{\bd})^{\dagger}~:=~(\bar \l^{\ad},\l^\b)\  ,\quad \pi(\l^\a,\bar\l^{\bd})~:=~(-\l^\a,\bar\l^{\bd})\ ,\quad \bar\pi (\l^\a,\bar\l^{\bd})~:=~(\l^\a,-\bar\l^{\bd})\ ,\quad \tau(\L^{\una})~=~i\L^{\una}\ .\ee
It follows that $(\Phi_{\L}(Y))^\dagger \equiv  (\Phi_{\L}(Y))   =  \Phi_{-\L}(Y)=\Phi_\Lambda(-Y)$, and hence, using the bosonic projection and after relabeling of the integration variables, one has $(\Phi'(Y))^\dagger= \int \frac{ d^4 \L}{(2 \pi)^2} \  \Phi_{\L}(Y) \ (\widetilde\Phi'(\L))^{\dagger}     $  implying the twisted-adjoint reality condition
\be (\widetilde \Phi')^{\dagger}~=~\pi(\widetilde \Phi')\ .\ee
One can also show that in the minimal bosonic model, the $\tau$-condition implies that
\be \tau(\widetilde \Phi')~=~\pi(\widetilde \Phi')\ ,\ee
from which follows the weaker bosonic projection $\pi\bar\pi(\widetilde \Phi')=\widetilde \Phi'$
As for the $\Phi'$-expansion of the primed master fields in quasi-amplitude computations, one defines
\be
\widehat\Phi^{\prime(n)'}(\Phi'_1, \dots, \Phi'_n)  ~=~  \left( \prod_{i = 1}^{n} \int  \frac{ d^4 \L_i}{(2 \pi)^2} \widetilde \Phi'_i(\L_i )\right)  \widehat\Phi^{\prime(n)'}_{\L_1,\dots,\L_n}(Y,Z) \ ,
\ee
where $\widehat\Phi^{\prime(n)'}_{\L_1,\dots,\L_n}:=\widehat\Phi^{\prime(n)'}(\Phi_{\L_1},\dots, \Phi_{\L_n} )$, \emph{idem} $\widehat S'_\a$.
As a result, the quasi-amplitudes can be written as
\bea
{\cal I}_{n_0}^{(n)'}  &=&  \left( \prod_{i = 1}^{n} \int  \frac{ d^4 \L_i}{(2 \pi)^2} \widetilde \Phi'_i( \L_i ) \right) \sum_{\tiny \ba{c}m_1 + \dots + m_{n_0}  = n\\ {\rm perm.}\ \L_i\ea} \widehat{\rm Tr}\left[\left( \prod_{i=1}^{n_0} \widehat  \Phi^{\prime (m_i)'}_{\L_{m_{i-1}+1} \dots \L_{m_i} } \star \widehat{\bar\kappa}\right) \star \widehat \kappa \widehat{\bar \kappa} \right] \ , 
\eea
and 
\bea
{\cal I}_{n_0}^{\prime(n)'}|_{\rm reg}  &=&  \int \frac{ d^2 \m d^2 \mb }{(2 \pi)^4} \widetilde{\cal V}'_{n_0}(\m , \mb)   \left( \prod_{i = 1}^{n} \int  \frac{ d^4 \L_i}{(2 \pi)^2} \widetilde \Phi'_i( \L_i ) \right) \sum_{\tiny \ba{c}m_1 + \dots + m_{n_0+2}  = n\\ {\rm perm.}\ \L_i\ea} \widehat{\rm Tr}\left[ \left( \prod_{i=1}^{n_0} \widehat  \Phi^{\prime (m_i)'}_{\L_{m_{i-1}+1} \dots \L_{m_i} } \star \widehat{\bar\kappa}\right)\right.\nonumber\\
&&\left. ~\qquad  ~\star ~\left( e_{\star}^{ i \m \widehat S'}\right)^{ (m_{n_0+ 1})' }_{\L_{m_{n_0} + 1}    \dots  \L_{m_{n_0 + 1} }}  \star \left( e_{\star}^{ -  i \mb \widehat S'}\right)^{ (m_{n_0+ 2})' }_{\L_{m_{n_0 + 1} + 1}    \dots  \L_{m_{n_0 + 2} }}   \right] \ ,
\eea
where $ \left( e_{\star}^{ i \m\widehat S'}\right)^{ (0) } =e^{i \m z }$ and the sub-leading terms can be obtained using Baker-Hausdorff-Campbell formula; for further details, see Appendix \ref{Exponential}. 

To examine these expansions, we use the realization \eqref{star product} of the $\star$-product, for which one has the following lemmas:
\bea
e^{i \l y -  i\lb \yb} \star \widehat f(y,\yb,z,\zb) &=&e^{i \l y -  i\lb \yb}  \widehat f(y - \l , \yb + \lb, z + \l,\zb + \lb) \ , \\ [5pt]  
 \widehat f(y,\yb,z,\zb) \star e^{i \l  y -  i\lb \yb}  &=& e^{i \l y -  i\lb \yb}  \widehat f(y + \l, \yb - \lb, z + \l,\zb + \lb ) \ ,
\eea
\bea
e^{i \mu z -  i\mb \zb} \star f(y,\yb,z,\zb) &=& e^{i \mu z - i\mb \zb} f(y - \mu, \yb - \mb, z + \mu,\zb - \mb) \ , \label{emuzstarf1}\\ [5pt]  
 f(y,\yb,z,\zb) \star e^{i \mu z -  i\mb \zb}  &=& e^{i \mu z - i\mb \zb} f(y - \mu, \yb - \mb, z - \mu,\zb + \mb) \ ,\label{emuzstarf2} 
\eea
\bea \widehat \kappa\star\widehat f(y,\yb,z,\zb) &=&\widehat \kappa f(z,\yb,y,\zb) \ ,\quad \widehat f(y,\yb,z,\zb) \star \widehat \kappa~=~\widehat \kappa f(-z,\yb,-y,\zb)\ ,  \\ [5pt]  
\widehat {\bar \kappa}\star\widehat f(y,\yb,z,\zb) &=&\widehat {\bar \kappa} f(y,-\yb,z,-\zb) \ ,\quad \widehat f(y,\yb,z,\zb) \star \widehat {\bar \kappa}~=~\widehat \kappa f(y,\zb,z,\yb)\ .  \eea

\subsection{Quasi-Amplitudes in Twistor Space with Two Legs}

Leaving ${\cal I}^{\prime(2)'}_1$ for a separate analysis, 
the quasi-amplitudes that are second order in $\Phi'$ are given by 
\bea
\label{amplitude 2}
{\cal I}^{(2)'}_2(\Phi'_1 , \Phi'_2)&= &  \frac12 \sum_{{\rm perm.} }   \widehat{\rm Tr}\left[ \Phi^{\prime}_1 \star\pi( \Phi^{\prime}_2)\star \widehat \kappa  \widehat{\bar{\kappa}}    \right]  \nonumber   \\[5pt]
 &= &  \left( \prod_{i = 1}^{2}  \int \frac{d^4 \L_i} {(2 \pi)^{2} }   \widetilde \Phi'_i( \L_i ) \right) \ \frac12 \sum_{{\rm perm.}\,\L_i  }\widehat{\rm Tr}\left[ e^{i \L_1Y }\star\pi( e^{i \L_2Y } ) \star \widehat \kappa  \widehat{\bar{\kappa}}    \right] \nonumber  \\[5pt]
 &= & \left( \prod_{i = 1}^{2}  \int \frac{d^4 \L_i} {(2 \pi)^{2} }   \widetilde \Phi'_i( \L_i ) \right)  \frac12 \sum_{{\rm perm.}\,\L_i  } e^{ - i \l_1 \l_2  +  i \lb_1 \lb_2 }
 \eea
\bea
\label{amplitude 2'}
{\cal I }^{\prime(2)'}_2(\Phi'_1 , \Phi'_2) &=&~ \sum_{{\rm perm.} } \widehat{\rm Tr}\left[ \Phi^{\prime}_1 \star\pi( \Phi^{\prime}_2  )\star \widehat{\cal V}'_2(z, \bar z )  \right] \ \nonumber \\[5pt]
 &= &   \left( \prod_{i = 1}^{2}  \int \frac{d^4 \L_i} {(2 \pi)^{2} }   \widetilde \Phi'_i( \L_i ) \right)   \int  \frac{d^2 \m d^2 \mb}{(2 \pi)^4} \widetilde{\cal V}'_2(\mu,\mb) ~\frac12  \sum_{{\rm perm.}\,\L_i  } \widehat{\rm Tr}\left[ e^{i \L_1Y }\star\pi( e^{i \L_2Y } ) \star e^{i\m z  -  i \mb \zb} \right] \nonumber  \\[5pt]
 &= &      \left( \prod_{i = 1}^{2}  \int \frac{d^4 \L_i} {(2 \pi)^{2} }   \widetilde \Phi'_i( \L_i ) \right)  \widetilde{\cal V}'_2( 0 , 0 ) \d^2(\l_1  - \l_2) \d^2(\lb_1 + \lb_2)  \ .
\eea

\subsection{Quasi-Amplitudes in Twistor Space with Three Legs}\label{Amplitudes}

In what follows, we shall need 
\be
\left( e^{i \m \widehat S'} \right)^{(1)'} ~=~    - \frac{b}2 \ \m z \int^{1}_{0} dt \, \frac{e^{ 2 i \m (y - \l ) t } - 1}{ \m (y - \l )   } \, e^{i (y - \l  - \m (1 - \frac{1}{t}))( z + \m) t  \  - i \lb  \yb } \\ [5 pt]
\ee 
where we have used
\be
\label{Auno}
\widehat V^{ \prime (1)'} ~=~  -\frac{i}2 \int^{1}_{0} dt\, t \left[  b \, z_{\a}   e^{i (y - \l) z \  t - i \lb \yb } ~ dz^{\a} ~ +~  \bar b \, \zb_{\ad}  \ e^{i\l y - i (\yb + \lb) \zb \  t } ~ d\zb^{\ad} \right]   \  ,  
\ee
which obeys the twistor gauge, and 
\bea
\label{fdue}
\widehat \Phi^{ \ \prime (2)'}_{\L_1, \L_2} & = &  \left.  - b  \ (z \l_2)  \int^{1}_{0} dt\ dt'   \  t   \ e^{i y(z tt' + \l_2(  t  - 1) ) -i (\lb_1 +\lb_2) \yb } ~\sin{\left( \lb_1\lb_2 + (\l_2  - \l_1)( zt t' + \l_2 t) \right)} \right|_{{\rm symm} \,1\leftrightarrow 2 } \nonumber \\[5pt]
&&~+ ~ \pi({\rm h.c. }) \ ,
\eea
where the symmetrization has unit strength, and the hermitian conjugation and the $\pi$-map act on $(Y,Z,\L)$.  
Leaving ${\cal I}^{\prime(3)'}_1$ for a separate analysis, 
the quasi-amplitudes with three external legs are  
\bea
{\cal I}^{(3)'}_2(\Phi'_1,\Phi'_2,\Phi'_3)   & = &  \frac1{3!}\sum_{{\rm perm.} }   \widehat{\rm Tr}\left[ \left(\widehat \Phi^{\prime (2)'} (\Phi'_1,\Phi'_2) \star\pi( \Phi^{\prime}_3) + \Phi^{\prime }_1 \star\pi( \widehat \Phi^{\prime (2)'} (\Phi'_2,\Phi'_3)) \right) \star \widehat \kappa  \widehat{\bar{\kappa}}    \right]    \nonumber\\[5pt] 
&  = &     \left( \prod_{i = 1}^{3}  \int \frac{d^4 \L_i} {(2 \pi)^{2} }   \widetilde \Phi'_i( \L_i ) \right) \widetilde{\cI}_2^{(3)'} ( \L_1, \L_2, \L_3)|_{\rm 2-hom.} \nonumber \\[5pt]
&  = &  0 , 
\eea
as can be seen from
\bea
\widetilde {\cI}_2^{(3)'}( \L_1, \L_2, \L_3)|_{2 - {\rm hom}}  &  = & 2  b  \    \d^2(\lb_1+ \lb_2 + \lb_3)   \int_0^1 dt dt' \frac{t}{(1  - tt')^3}   \nonumber \\[5pt]
&&\frac{1}{3!}\sum_{{\rm perm.}\,\L_i} (\l_2)^2 e^{i \l_2 \l_3 \frac{1 - t}{1  - t t'}} \sin{\left( \lb_1 \lb_2  - \l_1 \l_2 \frac{t (1  - t')}{1  - t t'} \right)}   + {\rm h.c.}\nonumber \\[5pt]
 &  = &  0  \ , 
\eea
in accordance with the claim that ${\cal I}_{n_0}$ are receive no perturbative corrections; and  
\bea
\label{amplitude 3pm}
{\cal I}^{\prime(3)' }_3(\Phi'_1,\Phi'_2,\Phi'_3)|_{\rm reg} & = & \frac{1}{ 3!} \sum_{{\rm perm.} }  \widehat{\rm Tr}\left[ \Phi^{\prime }_1\star\pi( \Phi^{\prime}_2 )\star \Phi^{\prime}_3 \star  \widehat{\kappa}  \star   \widehat{\cal V}_3(\bar z)  \right]  \nonumber\\[5pt]
&  = & \left( \prod_{i = 1}^{3}  \int \frac{d^4 \L_i} {(2 \pi)^{2} }   \widetilde \Phi'_i( \L_i ) \right) \widetilde{\cI}^{ \prime(3)' }_{3}( \L_1, \L_2, \L_3)\ ,
\eea
where 
\bea
\widetilde {\cI}_3^{\prime(3)'}( \L_1, \L_2, \L_3)  
 &=&  \widetilde{\cal V}_3(0)  \   \left.e^{-i( \l_1\l_2 + \l_2 \l_3 + \l_3\l_1 ) + i \lb_1\lb_2 }  \ \delta^2(\lb_1 + \lb_2 + \lb_3)\right|_{{\rm symm.}\,1\leftrightarrow 2}     \ ;
\eea
and finally
\bea
\label{amplitude 3'}
{\cal I}^{\prime(3)'}_2(\Phi'_1,\Phi'_2,\Phi'_3)|_{\rm reg}   & = &  \frac1{3!}\sum_{{\rm perm.}}  \left( \widehat{\rm Tr}\left[ \Phi^{\prime }_1  \star\pi( \Phi^{\prime}_2)\star  \widehat{\cal V}_2(\widehat S^{\prime(1)'}(\Phi'_3), \bar z )  \ + {\rm h.c.}   \right]    ~+~ \right.\nonumber \\[5pt]
 &&~ + \left. \widehat{\rm Tr}\left[ \left(  \widehat \Phi^{\prime (2)'} (\Phi'_1, \Phi'_2) \star\pi( \Phi^{\prime}_3)  +  \Phi^{\prime }_1 \star\pi( \widehat \Phi^{\prime (2)'}) (\Phi'_2, \Phi'_3)  \right)   \star \widehat{\cal V}_2(z,\bar z)     \right] \right) \ \   \nonumber\\[5pt] 
 &  = &     \left( \prod_{i = 1}^{3}  \int \frac{d^4 \L_i} {(2 \pi)^{2} }   \widetilde \Phi'_i(\L_i ) \right) \left( \widetilde {\cal I}^{\prime(3)'}_2( \L_1, \L_2, \L_3)|_{\rm 1-{\rm hom} }  + \widetilde {\cal I}^{\prime(3)'}_2( \L_1, \L_2, \L_3)|_{\rm 2-{\rm hom} } \right) \nonumber \\[5pt] 
 &  = &     \left( \prod_{i = 1}^{3}  \int \frac{d^4 \L_i} {(2 \pi)^{2} }   \widetilde \Phi'_i( \L_i ) \right) \widetilde {\cal I}^{\prime(3)'}_2( \L_1, \L_2, \L_3)|_{\rm 1-{\rm hom} }   ,  \qquad  
 \eea
where 
\bea
\widetilde {\cI}_2^{\prime(3)'}( \L_1, \L_2, \L_3)|_{1 - {\rm hom}}   &  = &   -  \frac{i b }{ 2\times 3!}\sum_{{\rm perm.}\,\L_i}e^{ - i (\l_1 \l_2 + \l_2 \l_3 + \l_3 \l_1) + i \lb_1 \lb_2} \delta^2(\lb_1 + \lb_2 + \lb_3) \nonumber\\[5pt] 
&&  \int \frac{d^2\m}{ (2 \pi)^2}  \widetilde{\cal V}_2(\m ,  0 )  \,\m(\l_1  - \l_2)   \int_0^1 \frac{d t}{t^2} e^{i \frac{t  - 1}{t } (\l_1  - \l_2) \m }    + {\rm h.c.}\nonumber \\[5pt]
&  = &  \frac{b}{2}   \left( \int \frac{d^2\m}{ 2 \pi} \widetilde{\cal V}_2(\m ,  0 ) \right)  \left. e^{-i( \l_1\l_2 + \l_2 \l_3 + \l_3\l_1 ) + i \lb_1\lb_2 }  \ \delta^2(\lb_1 + \lb_2 + \lb_3) \right|_{{\rm symm.}\,1\leftrightarrow 2}\nonumber\\&&~+~ {\rm h.c.}  \qquad , 
\eea
and we have used
\bea
\widetilde {\cI}_2^{\prime(3)'}( \L_1, \L_2, \L_3)|_{2 - {\rm hom}}    &  = &   - \frac{2 b}{3!} \int \frac{d^2\m d^2\mb }{ (2 \pi)^2} \widetilde{\cal V}_2(\mu,  \mb ) \d^{2}(\lb_1+ \lb_2 + \lb_3) \d^2(\mb)  \int_0^1 dt dt' \frac{t}{(t t')^3} \nonumber \\[5pt]
 &&~\sum_{{\rm perm.}\,\L_i} (\l_2 \l_3 )  \ e^{i \m( \l_2\frac{1  - t }{t t'}  + \l_3 \frac{t t'  - 1}{t t'}) } \sin{\left( \lb_1 \lb_2  - (\l_2  - \l_1 )(\l_3  - \l_2) \right)}+ {\rm h.c.}  \nonumber  \\[5pt]
&=&~  \frac{2  b}{3!}   \int \frac{d^2\m }{ (2 \pi)^2}  \widetilde{\cal V}_2(\mu,  0)  \sum_{{\rm perm.}\,\L_i} \frac{\l_2 \l_3 }{(\m\l_2  - \m\l_3)\m \l_3} \delta^2(\lb_1 + \lb_2 + \lb_3)  \nonumber \\[5pt]
&&~ \sin{\left( \lb_1 \lb_2  - (\l_2  - \l_1 )(\l_3  - \l_2) \right)}  + {\rm h.c.}   \nonumber \\[5pt]
 &=&~    \left.  \pm \frac{b}{ \pi } \widetilde{\cal V}_2(0,0) \,\delta^2(\lb_1 + \lb_2 + \lb_3)  \sin{\left( \lb_1 \lb_2  - (\l_1\l_2  + \l_2 \l_3 + \l_3 \l_1) \right)}  \right|_{{\rm symm.} \,1\leftrightarrow 2}\nonumber\\ &&~+~ {\rm h.c.} \nonumber  \\[5pt]
 &=&   0 \ .
\eea
In the above, the homotopy integrals are evaluated using the analytic continuation
\bea
I(A , B ) &  := & \int^{1}_{0} dt dt'\frac{1 }{t^2 t'^3} e^{ - i \frac{1  - t t' }{t t'}  A + i \frac{1  - t }{t t'}  B} ~=~  e^{ i A} \int^{1}_{0}\frac{dt'}{ t'^2 } e^{ - i \frac{1}{t'} B } \int^{\infty}_{\frac{1}{t'}} d\t  e^{- i \t (A  - B)} \\[5pt]
 &=&  \frac{1}{i (A  - B) } \int^{1}_{0}\frac{dt'}{t'} e^{ - i \frac{1}{t'} A} ~=~   \frac{1}{  i (A  - B) } \int_{1}^{\infty} d\s  e^{ - i \s A } ~=~  \frac{1}{(A  - B) A } \ ,
 \eea
where it is required that ${\rm Re}[ i (A  - B) ] =  {\rm Im } [ A - B ]  >   0 $   and  $ {\rm Re}[ i A ]  =  {\rm  Im } [A ] > 0$,  and the integral over $(\mu, \mb)$ is performed using the change of coordinates
\be
\mu_\a  ~= ~ u (\l_3  - \l_2)_\a ~+~ v (\l_3)_\a \qquad \Rightarrow \quad  \int d^2 \mu  =  - \l_2 \l_3 \int^{\infty}_{-\infty} du ~ dv 
\ee
leading to
\bea
 \int \frac{d^2\m}{ 2 \pi}  \frac{\l_2 \l_3}{\m(\l_3  -  \l_2 ) \m \l_3}   \widetilde {\cal V}_{2}(\mu,   0 )  &=&\frac{1}{2 \pi}  \int^{\infty}_{-\infty} du \  dv \  \frac{1}{u   v}  \widetilde {\cal V}_{2}( u (\l_3  - \l_2) + v \l_3 ,   0 ) \\ [5pt]
 & = & \frac{i}{2} \int^{\infty}_{-\infty}   dv   \frac{1}{ v}  \widetilde {\cal V}_{2}( v \l_3 ,   0 ) ~=~  \mp \frac{\pi}{2} \widetilde {\cal V}_{2}( 0 ,   0 ) \nonumber 
\eea 
in which we have used the principal value prescription to regularize the indefinite integral, \emph{i.e.} $\int \frac{da}{a} f(a)  =  \pm i \pi f( 0 ) $ where the sign depends on whether the function $f$ falls off sufficiently fast in the upper or the lower half plane; thus we are making the latter assumption on ${\cal V}_2(\mu,0)$.

\section{Current Correlation Functions from Zero-Form Charges}

In this section, we start from the twistor space quasi-amplitudes obtained from zero-form charges in the previous section and attach external legs given by the unfolded bulk-to-boundary propagators constructed by Giombi and Yin in \cite{Giombi:2010vg}, as to construct two-point and three-point amplitudes in the sector of states localized at the boundary of $AdS_4$.

\subsection{Unfolded Bulk-to-Boundary Propagators}

In what follows, we let $\Phi'_{s_i}(x_0;x_i)$ denote the value at a base point $x_0\in AdS_4$ of the linearized Weyl zero-form of a spin-$s_i$ bulk-to-boundary propagator whose source is localized at the point $x_i$ at the boundary and is taken to be magnetic for $s_i>0$, as to correspond to Dirichlet boundary conditions, and corresponding to Neumann boundary conditions for $s_i=0${}\footnote{Scalar fields with Dirichlet boundary conditions need a separate treatment.
}.
Working in Poincar\'e coordinates $x^\mu=(\vec x,r)$ where
\be
ds^2|_{AdS_4}  = \frac{1}{r^2 }  (dr^2 + d\vec x^2 )\ ,
\ee
where $d\vec x^2$ denotes the Minkowski metric and $r=0$ is identified as the conformal boundary of $AdS_4$; introducing the real boundary polarization bi-spinors
\be
\boldsymbol{\varepsilon}_{\a\b}  ~:=~ 4 \chi_{\a} \chi_{\b}\ ,\quad  \bar{\boldsymbol{\varepsilon}}_{\ad\bd}  ~:=~ 4 \bar\chi_{\ad} \bar\chi_{\bd}\ ,\quad (\chi_\a)^\dagger~:=~\bar\chi_{\ad}\ ,\quad \chi~:=~\sigma^r \bar \chi\ ,
\ee
where the factor $4$ is a  normalization choice; and assuming that the induced action of the parity map $P$ on the component fields $\Phi_i(x;x_i,s_i)$ is given by 
\be P\left(\boldsymbol{x}_{\a\bd}; ({\boldsymbol {x}}_i)_{\a\bd},(\chi_i)_\a,(\bar \chi_i)_{\ad}\right)~=~\left(\bar{\boldsymbol{x}}_{\ad\b}; (\bar{{\boldsymbol{ x}}}_i)_{\ad\b},(\bar\chi_i)_{\ad},( \chi_i)_{\a}\right)\ ,\ee
where $\boldsymbol{v}_{\a\bd}:=v^\mu(\sigma_\mu)_{\a\bd}$, it has been shown in \cite{Giombi:2009wh} that
\be
\Phi'_{s_i}(r_0, \vec x_0; \vec x_i,\chi_i; Y) ~:=~ \frac{1}{2}  \frac{(K_{i})^{2s_1+1}}{(2r_0)^{s_i}} e^{i y \boldsymbol{\Sigma_i} \yb} 
\left( b\left[ y  \boldsymbol{x}_{0i} \bar{\boldsymbol{ \e}}_i  \bar{\boldsymbol
{ x}}_{0i} y \right]^{s_i} + b^{-1} \left[ \yb  \bar{\boldsymbol{ x}}_{0i} 
\boldsymbol{ \varepsilon}_i  \boldsymbol{ x}_{0i} \yb \right]^{s_i} \right)\ ,
\ee
where
\be
x^\mu_{0i} ~:=~ x^\mu_{0}-x^\mu_{i}\ ,\quad 
 K_{i} ~:=~  (x_{0i})^{-2} r_0 \ , \quad \boldsymbol{\Sigma}_i  ~:= ~  \s^{r} - 2   r _0 \check{ \boldsymbol{ x}}_{0i}  \ ,\quad \check{x}^\mu_{0i}~:=~ (x_{0i})^{-2} x^\mu_{0i}\ ,
\ee
obeys the linearized equations of motion at the base point $x_0$ and 
\be P(\Phi'_{s_i})~=~b^2 \Phi'_{s_i}\ .\ee
These Weyl zero-forms thus describe linearized configuration of the Type A and Type B models, respectively, for $b=1$ and $b=i$. 
The inclusion of scalar fields with Neumann boundary conditions in the Type B model as well as of scalar fields in both models with Dirichlet boundary conditions require separate treatments, to be presented in the context of zero-form charges elsewhere.
As observed in \cite{Giombi:2010vg}, 
the above Weyl zero-forms can be obtained from the generating function
\be
 \Phi'_i   ~:=~  \sum_{s=0,2,\dots} \frac{1}{(2s)!} \Phi'_{s}(r_0, \vec x_0; \vec x_i,\chi_i; Y)~=~ \frac{1}{2}K_i e^{i y \boldsymbol{\Sigma}_i \yb}  \sum_{\s_i , \bar \s_i =\pm 1} \left( b e^{i \bar \s_i \nb_i \boldsymbol{\Sigma}_i y}  +b^{-1}  e^{i  \s_i \nu_i \boldsymbol{\Sigma}_i \yb }   \right)
\ ,\ee
where
\be \quad \nu_i ~ :=~ \sqrt{2r_0}\boldsymbol{\Sigma}_i\check {\boldsymbol{x}}_{0i}  \chi_i \ ,\quad (\bar\nu_i)_{\ad}~=~((\nu_i)_{\a})^\dagger\ ,\ee
with $Y$-space Fourier transform
\be
\widetilde \Phi'_i~:=~\widetilde \Phi'_i(r_0, \vec x_0; \vec x_i,\chi_i; \L_i) ~=~  \frac{1}{2}K_i e^{  - i \l_i \boldsymbol{\Sigma}_i \lb_i}  \sum_{\s_i , \bar \s_i}  \left(b e^{i \s_i \l_i\nu_i} + b^{-1} e^{i \bar \s_i \nb_i \lb_i }   \right)\ .
\ee

\subsection{Two-Point Functions}

As shown in Appendix \ref{App:Ampl}, the quasi-amplitudes \eqref{amplitude 2} and \eqref{amplitude 2'} yield the two-point functions
\bea
{\cal I}_2^{(2)'}(\vec x_1,\chi_1; \vec x_2,\chi_2) &  = & \int  \frac{d^4\L_1}{(2 \pi)^2} \frac{d^4\L_2}{{(2 \pi)^2}} \  \widetilde \Phi'_1(r_0,\vec x_0;\vec x_{1},\chi_1 ; \L_1)  \widetilde \Phi'_2(r_0,\vec x_0;\vec x_{2},\chi_2; \L_2)    \widetilde{\cal I}^{(2)'}_{2} (\L_1, \L_2)  \nonumber \\ [5pt]   
& = &    \frac{2}{ \vec{x}_{12}^2  }   \left(  1 +b^2 \cos \left( P_{12}\right) \right) \ ,  
\eea
\bea
{\cal I}^{\prime(2)'}_2(\vec x_1,\chi_1; \vec x_2,\chi_2 ) &  = & \int  \frac{d^4\L_1}{(2 \pi)^2} \frac{d^4\L_2}{{(2 \pi)^2}} \  \tilde \Phi_1(r_0,\vec x_0;\vec x_{1},\chi_1; \L_1)  \tilde \Phi_{2}(r_0,\vec x_0;\vec x_{2},\chi_2; \L_2)     \widetilde{\cal I'}^{(2)'}_{2} (\L_1, \L_2)  \nonumber \\ [5pt]   
& = & \widetilde {\cal V}_2(0,0)   \frac{2}{ \vec{x}_{12}^2  }   \left(  b^2 + \cos \left( P_{12}\right) \right)   \ ,  
\eea
where the conformally invariant variable \cite{Giombi:2011rz} 
\be    P_{12} ~ := ~  \bar \chi_1 \check{ \boldsymbol { x}}_{21} \bar \s^r  \bar \chi_2 ~=~ - \nu_1 \frac{ 1  - \boldsymbol{\Sigma}_1 \overline{ \boldsymbol{\Sigma}}_2  }{\det{(1  - \boldsymbol{\Sigma}_1 \overline{ \boldsymbol{\Sigma}}_2)}  }  \nu_2 \ ,\label{defP12}\ee
and the following identities have been used:
\be 
\label{sigma identities}
 \boldsymbol{\Sigma}\overline{\boldsymbol{\Sigma}_{i}}  ~=~  1\ ,\qquad (1  - \boldsymbol{\Sigma}_{i}\overline{\boldsymbol{\Sigma}}_{j})^{ - 1}  ~=~  \frac{ 1  -  \boldsymbol{\Sigma}_{j}\overline{\boldsymbol{\Sigma}}_{i}}{\det{ (1  - \boldsymbol{\Sigma}_{j}\overline{\boldsymbol{\Sigma}}_{i}) }}\ ,
\ee
\be
 \det{  (\boldsymbol{\Sigma}_i - \boldsymbol{\Sigma}_j) } ~=~ \det{  (1  - \boldsymbol{\Sigma}_i  \overline{\boldsymbol{\Sigma}}_j)}  ~=~ \frac{ 4 r^2}{x_{0i}^2 x_{0j}^2}  x^2_{ij}  ~ =~ 4 K_i K_j  x^2_{ij}    
\ee
\be
  {\check{\bar{\boldsymbol{ x}}}}_i (1  -  \boldsymbol{\Sigma}_i \overline{\boldsymbol{\Sigma}}_j)  \check{\boldsymbol { x}}_j  ~=~   \frac{ 2 r_0}{x_{0i}^2 x_{0j}^2}  ( \bar{\bold { x}}_i   -  \bar{\bold {x}}_j ) \s^r \ ,  \qquad  \bold {\check x}_i ( \overline{\boldsymbol{\Sigma}}_i  - \overline{\boldsymbol{\Sigma}}_j)  \bold {\check x}_j  =   \frac{ 2 r_0}{x_{0i}^2 x_{0j}^2}  ( \bold {x}_i   -  \bold {x}_j ) \ .  
\ee
Indeed, the two-point functions are independent of the choice of the base point, and the spin-$s$ contribution, which is obtained by extracting the $(\chi_1)^{2s} (\chi_2)^{2s}  \sim (\boldsymbol{ \varepsilon}_1)^s (\boldsymbol{ \varepsilon}_2)^s$ component, as explained in \cite{Giombi:2011rz}, reproduces the structures of point-split two-point correlation functions in three-dimensional free conformal field theories, \emph{viz.} 
\bea
\left\langle J_s( \vec x_{1}, \chi_1 )  J_s(\vec x_{2}  ,  \chi_2) \right\rangle ~=~     \frac{C_s}{ (\vec x_{12})^2}  \  (P_{12})^{2s}\ .\label{twopoints}
\eea

\subsection{Three-Point Functions}

\paragraph{Parity Invariant Quasi-Amplitude}

In Section \ref{Amplitudes}, it was shown that the quasi-amplitudes \eqref{amplitude 3pm} and \eqref{amplitude 3'} are proportional to a single basic building block for  amplitudes with three external legs, \emph{viz.}
\be {\cal I}_3^{\prime(3)'}|_{\rm reg}~=~\widetilde{\cal V}_3(0) {\cal I}^{(3)'}\ ,\quad {\cal I}_2^{\prime(3)'}|_{\rm reg}~=~ \frac{b}2 \left(\int \frac{d^2\mu}{2\pi} \widetilde{\cal V}_2(\mu,0)\right) {\cal I}^{(3)'}+{\rm h.c.}\ ,\ee
where the complex building block
\be {\cal I}^{(3)'}(\Phi'_1,\Phi'_2,\Phi'_3)~:=~\prod_{i=1}^3 \left(\int   \frac{d^4\L_i}{(2 \pi)^2} \widetilde \Phi'_i  \right)\widetilde{\cal I}^{(3)'} (\L_1, \L_2, \L_3)\ ,  \ee
with
\be
 \widetilde{\cal I}^{(3)'}(\L_1, \L_2, \L_3) ~=~ \left.e^{-i( \l_1\l_2 + \l_2 \l_3 + \l_3\l_1 ) + i \lb_1\lb_2 }  \ \delta^2(\lb_1 + \lb_2 + \lb_3)\right|_{{\rm symm.}\,1\leftrightarrow 2}\ .
\ee
In order to identify the structures of point-split three-point correlation functions in three-dimensional free conformal field theories,
one may observe that as these are parity invariant, the corresponding zero-form charges must be parity invariant as well. Thus, from  
\be P\left({\cal I}^{(3)'}\right)~=~ b^2 \left({\cal I}^{(3)'}\right)^\dagger\ ,\ee
it follows that the relevant candidates are 
\bea
\left.{\rm Re}\left[{\cal I}^{(3)'}\right]\right|_{{\rm A-mod.}} &  = &{\rm Re}\left[\left. \prod_{i=1}^3 \left(\int   \frac{d^4\L_i}{(2 \pi)^2} \widetilde \Phi'_i  \right)\right|_{{\rm A-mod.}} \widetilde{\cal I}^{(3)'} (\L_1, \L_2, \L_3)\right]   \\ [5pt]   
& = & \frac{1}{8} K_{1}K_{2} K_{3}    \sum_{\tiny \ba{c}\s_1, \s_2 , \s_3\\ \bar \s_1 , \bar \s_2,  \bar \s_3\ea  }  {\rm Re} \Big[\left.\left( 1 2 3  + \bar 1 2 3 +  1 \bar 2 3 + 1 2 \bar 3 + \bar 1 \bar 2 3 +  1 \bar 2 \bar 3 + \bar 1 2 \bar 3 + \bar 1 \bar 2 \bar 3  \right)\right|_{{\rm A-mod.}}  \Big]  \nonumber  \  , \\
\left.{\rm Im}\left[{\cal I}^{(3)'}\right]\right|_{{\rm B-mod.}}&  = & {\rm Im}\left[\left. \prod_{i=1}^3 \left(\int   \frac{d^4\L_i}{(2 \pi)^2} \widetilde \Phi'_i  \right)\right|_{{\rm B-mod.}}   \widetilde{\cal I}^{(3)'} (\L_1, \L_2, \L_3)\right]   \\ [5pt]   
& = &\frac{1}{8} K_{1}K_{2} K_{3}     \sum_{\tiny \ba{c}\s_1, \s_2 , \s_3\\ \bar \s_1 , \bar \s_2,  \bar \s_3\ea  } {\rm Im} \Big[  \left. \left( 1 2 3  + \bar 1 2 3 +  1 \bar 2 3 + 1 2 \bar 3 + \bar 1 \bar 2 3 +  1 \bar 2 \bar 3 + \bar 1 2 \bar 3 + \bar 1 \bar 2 \bar 3  \right)\right|_{{\rm B-mod.}}   \Big] \nonumber  \  ,
\eea 
where the separate contributions are given by (see Appendix \ref{App:Ampl} for further notation)   
\bea
(1 2 3)&=&  b^3 \prod_{i  = 1,2,3} \left(  \int  \frac{d^4\L_i}{(2 \pi)^2}    e^{-i \l_i \boldsymbol{\Sigma}_i \lb_i} \right)   e^{i \s_{1} \l_1 \nu_1  + \s_{2} \l_2 \nu_2 + \s_{3} \l_3 \nu_3 } \widetilde{\cal I}^{(3)'} (\L_1, \L_2, \L_3)  \ , \\[5pt]
(\bar 1 2 3) &=& b  \prod_{i  = 1,2,3} \left(  \int  \frac{d^4\L_i}{(2 \pi)^2} e^{-i \l_i \boldsymbol{\Sigma}_i \lb_i} \right)   e^{i \bar  \s_{1} \lb_1 \nb_1  + \s_{2} \l_2 \nu_2 + \s_{3} \l_3 \nu_3 } \widetilde {\cal I}^{(3)'} (\L_1, \L_2, \L_3)  \ , \\[5pt]
(\bar 1 \bar 2 3)&=& b^{-1}\prod_{i  = 1,2,3} \left(  \int  \frac{d^4\L_i}{(2 \pi)^2}  e^{-i \l_i \boldsymbol{\Sigma}_i \lb_i} \right)   e^{i \bar  \s_{1} \lb_1 \nb_1  + \bar \s_{2} \lb_2 \nb_2 + \s_{3} \l_3 \nu_3 }\widetilde {\cal I}^{(3)'}(\L_1, \L_2, \L_3) \ , \\[5pt]
(\bar 1 \bar 2 \bar 3) &=& b^{-3}  \prod_{i  = 1,2,3} \left(  \int \frac{d^4\L_i}{(2 \pi)^2} e^{-i \l_i \boldsymbol{\Sigma}_i \lb_i} \right)   e^{i \bar  \s_{1} \lb_1 \nb_1  + \bar \s_{2} \lb_2 \nb_2 + \bar \s_{3} \lb_3 \nb_3 } \widetilde{\cal I}^{(3)'} (\L_1, \L_2, \L_3) \ , 
\eea
and\footnote{These relations require only the cyclic symmetry of $ \widetilde{\cal I}^{(3)'} (\L_1,\L_2,\L_3)$ and not its total symmetry.}
\be (1 \bar 2 3) ~=~  ( 1  2 \bar 3)~=~ ( \bar 1  2  3)\ ,\quad ( 1 \bar 2 \bar 3)~ =~ ( \bar 1  2 \bar 3)~ =~(\bar 1 \bar 2 3)\ .\ee

\paragraph{Gaussian Integration and Cross Ratios} The integrals can be performed as follows:
\bea
 b^{-3}(1 2 3) &  = &   \int \frac{d^2 \l_1 d^2 \lb_1 d^2 \l_2 d^2 \lb_2 d^2 \l_3 d^2 \lb_3 } {(2 \pi)^{6} }  ~e^{ - i \l_1 \boldsymbol{\Sigma}_1 \lb_1 - i \l_2 \boldsymbol{\Sigma}_2 \lb_2   - i  \l_3 \boldsymbol{\Sigma}_3 \lb_3} \nonumber \\[5pt]
&&  e^{i \s_{1} \l_1 \nu_1  + \s_{2} \l_2 \nu_2 + \s_{3} \l_3 \nu_3 }    \left.\left[e^{i   \lb_1\lb_2} \delta^2(\lb_1 +  \lb_2 + \lb_3)   e^{-i(  \l_1\l_2 +   \l_2 \l_3 +  \l_3\l_1 )} \right] \right|_{\L_1\leftrightarrow \L_2}  \nonumber  \\ [5pt]
&  = &  \Big[\int \frac{d^2\l_1 d^2 \l_2 d^2 \l_3 d^2 \lb_3 } {(2 \pi)^{4} } \d^2\left(   \lb_3 +  \overline{\boldsymbol{\Sigma}}_{1} \l_1  - \overline{\boldsymbol{\Sigma}}_{2}\l_2 \right)     e^{   i \l_1 \boldsymbol{\Sigma}_{1}  \lb_3 - i  \l_3 \boldsymbol{\Sigma}_3 \lb_3 }  \nonumber \\[5pt]
&& \left.\left. e^{i \s_{1} \l_1 \nu_1  + \s_{2} \l_2 \nu_2 + \s_{3} \l_3 \nu_3 }    e^{-i(  \l_1\l_2 +   \l_2 \l_3 +  \l_3\l_1 )} \right]\right|_{(\vec x_1,\chi_1)\leftrightarrow (\vec x_2,\chi_2) } \nonumber \\[5pt]
 &  = & \Big[ \int \frac{d^2\l_2 d^2 \l_3  } {(2 \pi)^{2} }   \d^2\left(  - (1  - \boldsymbol{\Sigma}_{1}\overline{\boldsymbol{\Sigma}}_{2} ) \l_2  +  (1  - \boldsymbol{\Sigma}_{1}\overline{\boldsymbol{\Sigma}}_{3}) \l_3 + \s_1 \nu_1  \right) \nonumber \\[5pt]
&& \left.\left. e^{ - i  \l_2 (1  - \boldsymbol{\Sigma}_{2}\boldsymbol{\Sigma}_{3})\l_3 }  e^{i \s_{2} \l_2 \nu_2 + i \s_{3} \l_3 \nu_3 } \right]\right|_{(\vec x_1,\chi_1)\leftrightarrow (\vec x_2,\chi_2) } \nonumber \\[5pt]
 &  = &  \Big[ \frac{1}{\det{(1  -  \boldsymbol{\Sigma}_{1}\overline{\boldsymbol{\Sigma}}_{2})}} \int \frac{ d^2 \l_3  } {(2 \pi)^{2} } e^{i   \l_3\left[ (1  -  \boldsymbol{\Sigma}_{3}\overline{\boldsymbol{\Sigma}}_{2}) (1  - \boldsymbol{\Sigma}_{2}\overline{\boldsymbol{\Sigma}}_{1})^{ - 1}  (1  - \boldsymbol{\Sigma}_{1}\overline{\boldsymbol{\Sigma}}_{3}) \right] \l_3}  \nonumber\\[5pt]
&& e^{  i \s_{1}  \l_3  (1  -  \boldsymbol{\Sigma}_{3}\overline{\boldsymbol{\Sigma}}_{2}) (1  -  \boldsymbol{\Sigma}_{1}\overline{\boldsymbol{\Sigma}}_{2})^{ - 1} \nu_1  }   e^{  i  \s_{3}  \l_3 \nu_3  }  e^{  i \s_2  \l_3    (1  -  \boldsymbol{\Sigma}_{3}\overline{\boldsymbol{\Sigma}}_{1}) (1  -  \boldsymbol{\Sigma}_{2}\overline{\boldsymbol{\Sigma}}_{1})^{ - 1} \nu_2 } e^{   i \s_1 \s_2 \nu_2 (1  - \boldsymbol{\Sigma}_{1}\overline{\boldsymbol{\Sigma}}_{2})^{ - 1} \nu_1}\left.\Big]\right|_{(\vec x_1,\chi_1)\leftrightarrow (\vec x_2,\chi_2) } \nonumber \\[5pt]
 &  = & \Big[ \frac{1}{\det{ (1  -  \boldsymbol{\Sigma}_{1}\overline{\boldsymbol{\Sigma}}_{2}) } } \frac{1}{(2 \pi)^2}  \frac{\pi}{ 4 \sqrt{\det{A} }  } e^{ \frac{i}{4} J^{T} A^{ - 1} J } e^{   -  i \s_1 \s_2 \  \nu_1 \frac{ (1  - \boldsymbol{\Sigma}_{1}\overline{\boldsymbol{\Sigma}}_{2}) }{\det{ (1  -  \boldsymbol{\Sigma}_{1}\overline{\boldsymbol{\Sigma}}_{2})  } } \nu_2} \left.\Big]\right|_{1\leftrightarrow 2 } \ ,\label{prefactor}
\eea
 where we have defined 
 \bea
 J^{T}   &=& \left[  -   \s_1  \nu_1 \frac{(1  -  \boldsymbol{\Sigma}_{1}\overline{\boldsymbol{\Sigma}}_{2})}{\det{(1  - \boldsymbol{\Sigma}_{2}\overline{\boldsymbol{\Sigma}}_{1})}}  (1  - \boldsymbol{\Sigma}_{2}\overline{\boldsymbol{\Sigma}}_{3})    - \s_2  \nu_2  \frac{(1  -  \boldsymbol{\Sigma}_{2}\overline{\boldsymbol{\Sigma}}_{1})}{\det{(1  - \boldsymbol{\Sigma}_{1}\overline{\boldsymbol{\Sigma}}_{2})}}  (1  -  \boldsymbol{\Sigma}_{1}\overline{\boldsymbol{\Sigma}}_{3})    - \s_3  \nu_3  \right] \ ,\quad \quad\\[5pt]
 A^{ - 1} & = &  \frac{(1  -  \boldsymbol{\Sigma}_{3}\overline{\boldsymbol{\Sigma}}_{1})}{\det{(1  - \boldsymbol{\Sigma}_{1}\overline{\boldsymbol{\Sigma}}_{3})}}(1  -  \boldsymbol{\Sigma}_{1}\overline{\boldsymbol{\Sigma}}_{2}) \frac{(1  - \boldsymbol{\Sigma}_{2}\overline{\boldsymbol{\Sigma}}_{3})}{\det{(1  -  \boldsymbol{\Sigma}_{3}\overline{\boldsymbol{\Sigma}}_{2})}} \ , \\[5pt]
 J   &=& \left[  -   \s_1  (1  - \boldsymbol{\Sigma}_{3}\overline{\boldsymbol{\Sigma}}_{2})  \frac{(1  -  \boldsymbol{\Sigma}_{2}\overline{\boldsymbol{\Sigma}}_{1})}{\det{(1  - \boldsymbol{\Sigma}_{2}\overline{\boldsymbol{\Sigma}}_{1})}}    \nu_1   - \s_2   (1  -  \boldsymbol{\Sigma}_{3}\overline{\boldsymbol{\Sigma}}_{1})  \frac{(1  -  \boldsymbol{\Sigma}_{1}\overline{\boldsymbol{\Sigma}}_{2})}{\det{(1  - \boldsymbol{\Sigma}_{2}\overline{\boldsymbol{\Sigma}}_{1})}}   \nu_2    - \s_3  \nu_3  \right] \  . \quad \quad 
\eea    
Simplifying the exponentials, one finds 
\bea
(1 2 3) &\propto& \left.b^3 \exp{\left[ \frac{i}{4} \left( Q_1 + Q_2 + Q_3 \right) + \frac{i}{2}\left( \s_1 \s_2 P_{12} + \s_2 \s_3 P_{23} +  \s_3 \s_1 P_{31} \right) \right]}\right|_{1\leftrightarrow 2 } \ , \\ [5pt]
(\bar 1 2 3) &\propto&   \left.b \exp{\left[ \frac{i}{4} \left( Q_1 + Q_2 + Q_3 \right) + \frac{i}{2}\left( - \bar \s_1 \s_2 P_{12} + \s_2 \s_3 P_{23} + \s_3 \bar \s_1 P_{31} \right) \right] }\right|_{1\leftrightarrow 2 } \ , \\ [5pt]
(\bar 1 \bar 2 3)  &\propto&\left.b^{-1}  \exp{\left[ \frac{i}{4} \left( Q_1 + Q_2 + Q_3 \right) + \frac{i}{2}\left( - \bar \s_1 \bar \s_2 P_{12} - \bar \s_2 \s_3 P_{23} + \s_3 \bar \s_1 P_{31} \right) \right]}\right|_{1\leftrightarrow 2 } \ , \quad \\ [5pt]
(\bar 1 \bar 2 \bar 3)  &\propto&  \left.b^{-3} \exp{\left[  \frac{i}{4} \left( Q_1 + Q_2 + Q_3 \right) + \frac{i}{2}\left( - \bar \s_1 \bar \s_2 P_{12} - \bar \s_2 \bar \s_3 P_{23}  -  \bar \s_3 \bar  \s_1 P_{31} \right) \right]}\right|_{1\leftrightarrow 2 }  \ ,
\eea
where the proportionality is up to the real determinant pre-factor in \eqref{prefactor}, and the conformally invariant variables of \cite{Giombi:2011rz} has been identified as follows:  
\bea
P_{ij}  &:=& \bar \chi_i \bold{\check x}_{ij} \s \bar \chi_{j}~=~ - \nu_i  \frac{ (1  - \boldsymbol{\Sigma}_{i}\overline{\boldsymbol{\Sigma}}_{j}) }{\det{ (1  -  \boldsymbol{\Sigma}_{i}\overline{\boldsymbol{\Sigma}}_{j})  } } \nu_j  \ , \\[5pt]
Q_i  &:=&     \bar \chi_i \bold{\check x}_{i (i+1)}  \bold{x}_{(i+1)(i-1)}   \bold{\check x}_{(i-1)i} \s \bar \chi_{i}~=~-  \nu_i  \frac{ (1  - \boldsymbol{\Sigma}_{i}\overline{\boldsymbol{\Sigma}}_{i +1}) }{\det{ (1  -  \boldsymbol{\Sigma}_{i}\overline{\boldsymbol{\Sigma}}_{i +1})  } } (1  - \boldsymbol{\Sigma}_{i+1}\overline{\boldsymbol{\Sigma}}_{i  - 1})  \frac{ (1  - \boldsymbol{\Sigma}_{i - 1}\overline{\boldsymbol{\Sigma}}_{i }) }{\det{ (1  -  \boldsymbol{\Sigma}_{i - 1}\overline{\boldsymbol{\Sigma}}_{i })  } }   \nu_i  \ ,\qquad  \qquad 
\eea 
using ${\boldsymbol{\Sigma}}_{i + 3}\equiv {\boldsymbol{\Sigma}}_{i} $. 

\paragraph{A Model and Free Scalars}

In  the A model, the sum 
\bea
&&  \sum_{\tiny \ba{c}\s_1, \s_2 , \s_3\\ \bar \s_1 , \bar \s_2,  \bar \s_3\ea  }  \left(  (1 2 3)  + (\bar 1 \bar 2 \bar 3)\right)|_{\rm A-mod.}  \\ [5pt]
&  \propto & \left.16 \sum_{\s_1, \s_2}  e^{\frac{i}{4} \left( Q_1 + Q_2 + Q_3 \right) }    \cos \left(\frac{1}{2} \s_2 P_{12}  +  \frac{1}{2}\s_3 P_{31} \right) \left( e^{ \frac{i}{2} \s_2 \s_3 P_{23} }  + e^{- \frac{  i}{2} \s_2 \s_3 P_{23} }  \right)\right|_{1\leftrightarrow 2 }\nonumber  \\[5pt]
&   = & \left.64  \ e^{\frac{i}{4} \left( Q_1 + Q_2 + Q_3 \right) }   \left[  \cos\left( \frac{1}{2} P_{23} \right)   \cos \left(\frac{1}{2} P_{12}  +  \frac{1}{2} P_{31} \right)  +  \cos\left( \frac{1}{2} P_{23} \right)   \cos \left(\frac{1}{2} P_{12}  -  \frac{1}{2} P_{31} \right)  \right]\right|_{1\leftrightarrow 2 } \nonumber  \\[5pt]
&  =  & \left.64 \  e^{\frac{i}{4} \left( Q_1 + Q_2 + Q_3 \right) }   \cos\left( \frac{1}{2} P_{23} \right)   \cos \left(\frac{1}{2} P_{12}\right) \cos \left( \frac{1}{2} P_{31} \right)\right|_{1\leftrightarrow 2 }  \ \nonumber \\[5pt]
&  =  & 64 \  \cos\left[\frac{1}{4} \left( Q_1 + Q_2 + Q_3 \right) \right]   \cos\left( \frac{1}{2} P_{23} \right)   \cos \left(\frac{1}{2} P_{12}\right) \cos \left( \frac{1}{2} P_{31} \right)\ \ , 
\eea
\emph{idem.} the other six contributions $((\bar 1 2 3)+ (\bar 1 \bar 2 3)+ {\rm cyclic})|_{\rm A-mod.} $, where the last equality shows that the parity-odd imaginary part of the quasi-amplitude indeed vanishes, \emph{i.e.}
\be
{\rm Re}\left[\left.{\cal I}^{(3)'}\right]\right|_{{\rm A-mod.}}  ~=~   \left.{\cal I}^{(3)'}\right|_{{\rm A-mod.}}\ .\ee
Attaching the pre-factor containing the determinants, the remaining dependence on the base point drops out\footnote{From the definition of $K_1 , K_2 , K_3$ and the identities \eqref{sigma identities}, one has
$$
 \frac{K_1 K_2 K_3}{8} \frac{1}{\det{(1  -  \Sigma_{1}\Sigma_{2})}} \frac{1}{(2 \pi)^2} \frac{\pi}{\sqrt{\det{A} } }  ~ =~    \frac{1}{8} \frac{r_0}{x_{1}^{2}}  \frac{r_0}{x_{2}^{2}}  \frac{r_0}{x_{3}^{2}} \frac{1}{(2 \pi)^2}  \frac{\pi\sqrt{x^{4}_{1} x^{4}_{2} x^{4}_{3} }}{\sqrt{64 r^{6}_0 (\vec x_{12})^2(\vec x_{23})^{2} (\vec x_{31})^{2}  } }~  = ~  \frac{1}{64}   \frac{1}{(4 \pi)} \frac{1}{|\vec{x}_{12}| |\vec{x}_{23}| | \vec{x}_{31}| } \   \ .  
 $$
}, and one arrives at 
 \be
 \left.{\cal I}^{(3)'}\right|_{{\rm A-mod.}}~=~\frac{1}{4 \pi} \frac{1}{|\vec{x}_{12}||\vec{x}_{23}||\vec{x}_{31}|}   \cos \left[ \frac{1}{4} \left( Q_1 + Q_2 + Q_3 \right) \right]   \cos\left( \frac{1}{2} P_{23} \right)   \cos \left(\frac{1}{2} P_{12}\right) \cos \left( \frac{1}{2} P_{31} \right) \  ,  \label{threepointsA}
\ee
where the right-hand side can be identified as the generating function for point-split three-current correlation functions in free scalar conformal field theory in three dimensions \cite{Giombi:2011rz}. 

\paragraph{B Model and Free Fermions}

In the B model the summation over $\s_i$ and $\bar \s_i$ instead yields 
\bea
&&  \sum_{\tiny \ba{c}\s_1, \s_2 , \s_3\\ \bar \s_1 , \bar \s_2,  \bar \s_3\ea  }  \left(  (1 2 3)  + (\bar 1 \bar 2 \bar 3) \right) |_{\rm B-mod.}  \\ [5pt]
&  \propto & - \left.16i \sum_{\s_1, \s_2}  e^{\frac{i}{4} \left( Q_1 + Q_2 + Q_3 \right) }    \cos \left(\frac{1}{2} \s_2 P_{12}  +  \frac{1}{2}\s_3 P_{31} \right) \left( e^{ \frac{i}{2} \s_2 \s_3 P_{23} }   -  e^{ -\frac{  i}{2} \s_2 \s_3 P_{23} }  \right)\right|_{1\leftrightarrow 2 }\nonumber  \\[5pt]
&   = & \left.64 \ e^{\frac{i}{4} \left( Q_1 + Q_2 + Q_3 \right) }   \left[  \sin\left( \frac{1}{2} P_{23} \right)   \cos \left(\frac{1}{2} P_{12}  +  \frac{1}{2} P_{31} \right)   -   \sin\left( \frac{1}{2} P_{23} \right)   \cos \left(\frac{1}{2} P_{12}  -  \frac{1}{2} P_{31} \right)  \right] \right|_{1\leftrightarrow 2 }\nonumber  \\[5pt]
&  =  &  \left.64  \ e^{\frac{i}{4} \left( Q_1 + Q_2 + Q_3 \right) }   \sin\left( \frac{1}{2} P_{23} \right)   \sin \left(\frac{1}{2} P_{12}\right) \sin \left( \frac{1}{2} P_{31} \right) \right|_{1\leftrightarrow 2 } \ \nonumber \\[5pt]
&  =  &  64 i \sin\left[\frac{1}{4} \left( Q_1 + Q_2 + Q_3 \right) \right]   \sin\left( \frac{1}{2} P_{23} \right)   \sin \left(\frac{1}{2} P_{12}\right) \sin \left( \frac{1}{2} P_{31} \right) \ \nonumber \  ,
\eea
with similar results for the remaining contributions. One obtains 
\be   {\rm Im}\left[\left.{\cal I}^{(3)'}\right]\right|_{{\rm B-mod.}}~=~\left.-i {\cal I}^{(3)'}\right|_{{\rm B-mod.}}\ ,\ee
where
 \be
\left.-i{\cal I}^{(3)'}\right|_{{\rm B-mod.}} ~=~   \frac{1}{4 \pi}   \frac{1}{|\vec{x}_{12}||\vec{x}_{23}||\vec{x}_{31}|}   \sin \left[ \frac{1}{4} \left( Q_1 + Q_2 + Q_3 \right) \right]   \sin\left( \frac{1}{2} P_{23} \right)   \sin \left(\frac{1}{2} P_{12}\right) \sin \left( \frac{1}{2} P_{31} \right)\ ,\label{threepointsB}
\ee
where the right-hand side reproduces the point-split three-current correlation functions in free fermion conformal field theory in three dimensions \cite{ Giombi:2011rz}. 

\section{Towards an On-Shell Action in Twistor Space}\label{Sec:action}

\paragraph{On-Shell Assumptions}

In order to proceed, we suppose that the results reported so far admit a natural generalization to any number of points, \emph{i.e.} given a definite value of $b$ and working with suitable unfolded bulk-to-boundary propagators, and letting ${\cal C}^{(n)}|_{\rm split}$ denote the canonically normalized point-split $n$-current correlation functions in the corresponding three-dimensional conformal field theories, we make 
\begin{itemize}
\item[] \emph{Assumption 1:} ${\cal I}_{n_0}={\cal I}_{n_0}^{(n_0)}$ generates point-split $n_0$-current correlation functions, \emph{viz.}
\be {\cal I}^{(n_0)}_{n_0}~=~ {\cal N}_{n_0} {\cal C}^{(n_0)}|_{\rm split}\ ,\qquad n_0~=~2,4,\dots\ ;\ee
\item[] \emph{Assumption 2:} ${\cal I}'_{n_0}|_{\rm reg}=\sum_{n\geqslant n_0}{\cal I}_{n_0}^{\prime(n)}|_{\rm reg}$ generate point-split $n$-current correlation functions together with $n$-point functions that are supported in coincidence limits, \emph{viz.}
\be {\cal I}^{\prime(n)}_{n_0}|_{\rm reg}~=~ {\cal N}^{\prime(n)}_{n_0}{\cal C}^{(n)}|_{\rm split}+ {\cal C}_{n_0}^{(n)}|_{\rm coin}\ ,\label{coin}\ee
where the on-shell coupling ${\cal N}_{n_0}^{\prime(n)}$ is given by a moment of $\widetilde{\cal V}_{n_0}$ or its Fourier transform, and ${\cal C}_{n_0}^{(n)}|_{\rm coin}$ is projected by $\delta^3(\vec x_{ij})$ -functions.
\end{itemize} 

\paragraph{Off-Shell Assumptions}

We proceed by stating assumptions that are necessary in order for the zero-form charges to arise as the on-shell values of meaningful deformations of the generalized Hamiltonian action for Vasiliev's four-dimensional higher spin gravities constructed in \cite{Boulanger:2011dd}, namely that they admit off-shell resolutions as topological vertex operators, that is, functionals defined off shell whose total variation vanishes on shell \cite{Sezgin:2011hq} (see also \cite{Boulanger:2012bj}).
Drawing on the similarity between the twistor space regulators $\widehat{\cal V}_{n_0}$ and ordinary Wilson loops, we are led to make 
\begin{itemize}
\item[] \emph{Assumption 3:} For each $n_0$, the requirement that ${\cal I}_{n_0}' $ admits an off-shell extension as a topological vertex operators in the sense of \cite{Sezgin:2011hq}\footnote{For a more recent discussion, see also \cite{Boulanger:2012bj}.} determines $\widehat {\cal V}_{n_0}$ up to an overall normalization ${\cal N}'_{n_0}$;
 \item[] \emph{Assumption 4:} The form of the topological vertex operator is compatible with finiteness of the on-shell couplings, \emph{viz.}
\be {\cal N}^{\prime(n)}_{n_0}~\equiv~{\cal N}'_{n_0}{\cal V}^{(n)}_{n_0}\ ,\ee
where ${\cal V}^{(n)}_{n_0}$ are uniquely determined by Assumption 3. 
\end{itemize} 

\paragraph{Free Energy} The on-shell action is defined by
\be {\cal I}_\mu~=~\sum_{n_0}{\cal I}'_{n_0}|_{\rm reg}+\sum_{n_0} \mu_{n_0} {\cal I}_{n_0}\ ,\label{action}\ee
where $\mu_{n_0}$ play the r\^ole of generalized chemical potentials  \cite{Boulanger:2012bj} while the normalizations ${\cal N}'_{n_0}$ are fixed by demanding that the point-split $n$-point amplitudes are given by ${\cal C}^{(n)}|_{\rm split}$, \emph{i.e.} 
\be {\cal N}'_{n_0}{\cal V}_{n_0}^{(n)}+\mu_n {\cal N}_n ~=~1\ ,\ee
which is tantamount to requiring on-shell cluster decomposition.
By our hypotheses, the on-shell action a) reproduces the point-split multi-current correlation functions; b) contains sub-leading corrections corresponding to contributions to multi-current correlation functions on hyper surfaces where points coincide; and c) yields finite free energies for nontrivial classical solutions in topologically trivial situations. 

Concerning (c), the zero-form charges ${\cal I}_{n_0}$ are well-defined on a number of exact solutions\footnote{In fact, in all solutions that have appeared in the literature so far (for a brief review, see for example \cite{Iazeolla:2011cb}), there exists a gauge in which the Weyl zero-form is independent of $Z^{\una}$, \emph{i.e.} $\widehat \Phi=\Phi$, and $\Phi$ consists either of only the scalar field or of infinite towers of fields. The Ansatz of \cite{Iazeolla:2011cb} can easily be modified, however, as to yield exact solutions in which $\widehat \Phi=\Phi$ consists of a single spin-$s$ field propagating in its lowest-weight space $D(s+1,(s))$.}
provided that they are presented as integrals of master-field constructs over twistor space rather than as strongly coupled derivative expansions in terms of component fields in spacetime\footnote{In the case of spherically symmetric solutions, twistor space is instrumental not only for finiteness of zero-form charges but also for resolving singularities at the origin \cite{Iazeolla:2011cb}; at this point the Weyl zero-form, given in the normal order corresponding to \eqref{star product}, approaches a distribution on ${\cal Y}\times {\cal Z}$ given by derivatives of $\delta^2(y_\a-i(\s^0)_{\a}{}^{\ad}\yb_{\ad})$, which implies that each separate spin-$s$ Weyl tensor diverges, as $r^{-s-1}$, while the zero-form charges ${\cal I}_{n_0}$ remain finite. }.
The fact that the zero-form charges are finite provided one first sums over all Lorentz spins and then integrates over twistor space manifests itself in the soft behaviors of \eqref{twopoints}, \eqref{threepointsA} and \eqref{threepointsB} in coincidence limits; indeed, if an exact solution admits a perturbative description starting from a source
\be \Phi'(r_0,\vec x_0)~=~\int d^3\vec x d^2\chi \phi(\vec x,\chi) \Phi'(r_0,\vec x_0; \vec x,\chi)\ ,\label{sources}\ee
where thus $\phi(\vec x,\chi)$ is a finite classical source,
then  
\be {\cal I}_{2}~=~\int d^3\vec x_1 d^2\chi_1\int d^3\vec x_2 d^2\chi_2 |\vec x_1-\vec x_2|^{-2}\left(1+b^2 \cos(P_{12})\right) \phi(\vec x_1,\chi_1)\phi(\vec x_2,\chi_2)\ ,\ee
which is free from singularities as $P_{12}$ is a real variable which means that the integrand is bounded (and in fact oscillates rapidly) in the limit $\sim |\vec x_1-\vec x_2|^{-2}$. 

\paragraph{Holographic Correspondence}

A natural interpretation of sub-leading corrections is to identify ${\cal I}_0\equiv {\cal I}_{\mu}|_{\mu_{n_0}=0}$ with the generating functional of a suitably regularized three-dimensional conformal field theory\footnote{It would be interesting to examine in more detail to what extent the freedom in redefining $\Phi'$ without affecting the value of ${\cal I}_{\mu}$, for example by modifying the twistor gauge condition, corresponds to the freedom in choosing regularization scheme for computing the counter terms \cite{Douglas:2010rc}.}
:
\be \exp(i{\cal I}_0(\Phi'))~=~\left.\left\langle \exp\left( i\int d\vec x \left(\sum_{s} J_{s}(\vec x) \phi_{s}(\vec x)+\sum_{\ell} \Gamma_\ell (\vec x) \Sigma_\ell(\vec x)\right)\right)\right\rangle\right|_{\rm reg}\ ,\label{holoprop}\ee
where $J_s$ denote the bilinear higher-spin currents with finite sources $\phi_{s}$ defined by \eqref{sources}, and $\{\Gamma_\ell\}$ is a complete set of operators, containing the identity, the currents and normal-ordered products of the currents, with finite sources $\Sigma_\ell=\sum_{n\geqslant 2}\Sigma^{(n)}(\underbrace{\phi,\dots,\phi}_{\tiny\mbox{$n$ times}})$ that are supported only in coincidence limits such that taking $n$ functional derivatives with respect to $\phi_s$ yields the contact terms ${\cal C}^{(n)}|_{\rm coin}$ in \eqref{coin}.
Thus, from the point-of-view of holographic correspondence, the leading contributions to the zero-form charges, which are fixed entirely by kinematics without referring to the non-linearities of the Vasiliev system, correspond to point-split contributions to the multi-current correlation functions, while the sub-leading corrections, which refer to the nonlinearities in the Vasiliev system, correspond to contributions in coincidence limits,
\emph{i.e.} contact terms (for example, see \cite{Closset:2012vp})\footnote{Drawing on the recent proposal in \cite{Vasiliev:2012vf}, (see also \cite{Nilsson:2012ky,Gran:2012mg} for a related proposal in the context of gauged supergravities), it would be interesting to seek a constructive approach towards regularizing three-dimensional conformal field theories via coupling to topological conformal higher spin gravity fields with classical expectation values interpreted as finite sources.}.

\section{Conclusions}

We conclude by collecting some remarks, summarizing our results and the procedure proposed above for obtaining the on-shell action, after which we present a brief outlook.

\subsection{Select Remarks}

We would like to make the following comments and remarks:

\begin{itemize}
\item \emph{Higher-point functions and cyclic structures:} 
While the quasi-amplitudes are manifestly cyclically invariant, we expect that the correspondence between traces of unfolded bulk-to-boundary propagators over ${\cal Y}$-space and multi-current correlation functions in theories of free scalar and fermions hold already at the level of cyclically invariant structures, \emph{viz.} 
%

\bea \mbox{$n_0$ even}&:&  \left.{\rm Tr}\left[\prod_{i=1}^{n_0}{}^\star \left(\Phi'_i(
\vec x_i,\chi_i;Y) \star \bar\kappa_{\yb }\right)\right]\right|_{\rm A-mod.}\nonumber\\[5pt]&&~\sim~\left.{\rm 
Tr}\left[\left(\prod_{i=1}^{n_0}{}^\star \Phi'_i(\vec x_i,\chi_i;Y) \star \bar\kappa_{\yb }\right)
\star\kappa_y\bar\kappa_{\yb}\right]\right|_{\rm A-mod.}\nonumber\\[5pt]&&~\sim~{\cal 
C}^{(n)}_{{\rm scalar}}(\vec x_1,\chi_1;\cdots;\vec x_n,\chi_n)|_{\rm cyclic}\ ,\\[5pt] \label{propamodel}\mbox
{$n_0$ odd}&:&  \left.{\rm Tr}\left[\prod_{i=1}^{n_0}{}^\star \Phi'_i(\vec x_i,\chi_i;Y) \star \bar
\kappa_{\yb }\right]\right|_{\rm A-mod.}~\sim~{\cal C}^{(n)}_{{\rm scalar}}(\vec x_1,\chi_1;\cdots;\vec x_n,\chi_n)|_
{\rm cyclic}\ ,\qquad\label{propbmodel}
\eea

\emph{idem} for the $B$-model and free fermions,
where ${\rm Tr}[f(y,\yb)]\equiv \int\frac{d^4Y}{(2\pi)^2}f(y,\yb)$ using symbol calculus and  $\kappa_{y}=2\pi '\delta^2(y)$ using Weyl order \emph{idem} $\bar\kappa_{\yb}$, and ${\cal C}^{(n)}_{{\rm scalar}}|_{\rm cyclic}$ and ${\cal C}^{(n)}_{{\rm fermion}}|_{\rm cyclic}$ are calculated by contracting pairs of free fields at $(\vec x_i,\chi_i)$ and $(\vec x_{i+1},\chi_{i+1})$ (with $(\vec x_{n+1},\chi_{n+1})\equiv (\vec x_1,\chi_1)$).
\item \emph{Parity violating conformal blocks:} 
As the parity violating terms in Vasiliev's equations only affect sub-leading orders, and as we have seen that conformal blocks can appear both at leading and sub-leading orders, we expect that\footnote{
The linearized Vasiliev's equations with general $b$ do not explicitly violate parity nor higher-spin Killing symmetries. However, both parity and rigid higher-spin symmetry can be broken in the linearized approximation by imposing suitable boundary conditions \cite{Chang:2012kt}.
} the parity violating structure for the three-point functions \cite{Giombi:2011rz} arises in ${\cal I}_1^{(3)}$.
\item \emph{Free energy as function of asymptotic charges:} %
In \cite{Iazeolla:2011cb}, several families of exact solutions have been constructed in 
which the Weyl zero-form is of the form $\widehat \Phi'=\sum_{\vec n} \nu_{\vec n} T_
{\vec n}(Y)$, where $\nu_{\vec n}$ are complex numbers and $T_{\vec n}$ are diagonal 
elements in bases for the twisted-adjoint representations associated to various types of 
boundary conditions labelled by two charges in the complexified Cartan subalgebra of 
$so(3,2)$. In particular, using the AdS energy and a spin yields a standard compact basis 
and the solutions are stationary and asymptotic to $AdS_4$. As the $T_{\vec n}$ form 
an algebra and their supertraces are finite, it follows that ${\cal I}_{n_0}\sim \sum_{\vec n} 
(\nu_{\vec n})^{n_0} c_{n_0,\vec n}$ where $c_{n_0,\vec n}$ are constants related to the 
supertrace of $T_{\vec n}$. On the other hand, by introducing gauge functions one can 
map the $\nu_{\vec n}$ to sets of asymptotically defined charges ${\cal M}_s=\sum_{\vec 
n} {\cal M}_s^{\vec n} \nu_{\vec n}+\sum_{n\geqslant 2}{\cal M}_s^{(n)}$ labelled by 
Lorentz spins $s$ and where the higher order corrections ${\cal M}_s^{(n)}$, of $n$th 
order in $\nu_{\vec n}$, arise upon imposing the twistor gauge \eq{twistorgauge}.  
Assuming that there exists an inverse map, $\nu_{\vec n}=\sum_s {\cal M}_{\vec n}^s {\cal 
M}_s +\sum_{n\geqslant 2} \nu_{\vec n}^{(n)}$, which is by no means clear in general as 
the two sets of indices $s$ and $\vec n$ refer to different bases of an infinite-
dimensional representation space,  one has ${\cal I}_{n_0}={\cal I}_{n_0}({\cal M}_s)$ 
that one may think of as generalized Casimir invariants, which are indeed the natural 
basic building blocks for constructing free energies and entropies in ensembles of 
localizable classical solutions\footnote{In higher spin gravity, the fact that four-
dimensional spacetime is acted upon by both spin-$2$ symmetries as well as higher spin 
symmetries means that there is no invariant meaning to an asymptotic charge ${\cal M}_s
$ for a given spin $s$.}.
\item \emph{Compact basis and sewing operation:} 
In the compact basis, where states are labeled by anti-de Sitter energies and $so(3)$ 
spins, we expect the zero-form charges to match quantities obtained in the formulation of the free theories on 
$S^2\times S^1$ \cite{Bergshoeff:1988jm}. As the compact basis is discrete, it would be interesting to examine whether it can facilitate the gluing together of external legs on amplitudes, as to add radiative bulk corrections or, correspondingly, as to emulate the sewing operation proposed in \cite{Sezgin:2002rt} in order to add $1/N$-corrections to the boundary theory associated to formulating it on other topologies than $S^2$.
\item \emph{Soliton sector:} 
Linearized higher spin gravity contains smooth solutions that are static and rotationally invariant, and one may argue \cite{Iazeolla:2008ix} that they admit a perturbative completion. Although these solution-like solutions do not fall off fast enough at large radii as to have finite canonical Killing energy in the sense of \cite{Breitenlohner:1982jf}, they do nonetheless give rise to unitarizable representations of the higher spin algebra \cite{Iazeolla:2008ix}.
It would thus be interesting to use the twistor-space method to examine whether they are localizable at the level of suitable quasi-amplitudes, and furthermore whether they have a finite regularized free energy in the sense proposed in this paper.
\item \emph{Higher dimensions:} 
In higher-dimensional bosonic theories based on vector oscillators \cite{Vasiliev:2004cp}, the zero-form charges are given by traces with insertions of the quasi-
projector $\widehat M$, \emph{viz.} $\widehat{\rm Tr}[\widehat M\star (\widehat \Phi\star 
\widehat \kappa)^n\star \exp^{i\mu\widehat S}_\star]$ where $\widehat M$; see also \cite{Engquist:2005yt}. 
As in four dimensions, the leading orders are determined entirely by 
kinematics, and we expect that these correspond to the point-split multi-current 
correlation functions in the free scalar theory; these kinematic considerations can also be 
generalized to the undeformed five-dimensional and seven-dimensional higher spin 
gravities based on twistor oscillators constructed in \cite{Sezgin:2001zs,Sezgin:2001ij}, 
including the supersymmetric models given in \cite{Sezgin:2001yf} and \cite{Sezgin:2002rt} corresponding, respectively, to the free superconformal ${\cal N}=4_4$ and ${\cal N}=(2,0)_6$ theories. 
As for sub-leading terms, the regularization of the zero-
form charges in higher dimensions ought to be more subtle than that in four dimensions, tied to the expected non-triviality of renormalization group flows and conformal fixed points in higher-dimensional field theory\footnote{Indeed, while the the four-dimensional Vasiliev models in twistor space admit a intricate deformations  \cite{Vasiliev:1990en,Sezgin:2011hq}, suggesting that there exist a vast conformal field theory landscape in three dimensions, the analog can be removed by means of perturbatively defined field redefinitions from the higher-dimensional Vasiliev models based on vector oscillators. }.
\item \emph{Quasi-actions and spacetime pictures:} The moduli contained in the boundary values of the gauge 
functions, which are nontrivial only in topologically broken phases of the theory, and 
which one may think of as generalized Brown--Henneaux boundary gravitons, are measured by quasi-actions depending on a soldering one-form; see Eq. \eqref{quasi-action}. In \cite{Sezgin:2011hq}  it was found that if the structure group is taken to be 
the product of the group of manifest Lorentz symmetries and the group of higher spin 
transformations generated by $\pi$-even gauge parameters\footnote{These choices are by no means unique. For example, in the structure group one may replace the group of $\pi$-even elements by the group of elements that are holomorphic in ${\cal Y}\times {\cal Z}$, as pointed out by Vasiliev, or even the trivial group, in which case the structure group would be given simply by the group of manifest Lorentz symmetries.}, then there exists one complex cohomologically nontrivial $p$-form on shell for each $p=2,4,\dots$.
Moreover, these were shown to admit off-shell extensions as topological vertex operators, such that one may think of them as order parameters for a metric phase with soldering one form $\widehat E=\frac12(1-\pi)\widehat W$. 
In odd form-degree, on the other hand, there exists a large number of formally on-shell exact forms. One may thus speculate that a suitable subset of the latter are actually on-shell restrictions of topological vertex operators that develop singularities, as to become cohomologically nontrivial, in sectors of localizable states, such as for example the aforementioned bi-axially symmetric solutions.
If so, in each even dimension, there would exist a spacetime picture, with total on-shell action given by the sum of the closed $p$-form, functioning as the generating function of boundary correlation functions, plus the tower of $(p-1)$-form charges coupled to chemical potentials.
In this scenario, the $(p-1)$-form charges would thus be given by two surface terms: one at the center of the solution, representing the Weyl zero-form moduli, and one at infinity, mixing these modulo with those of the generalized boundary gravitons.
Thus, as each charge activates only one chemical potential, to be associated with the Weyl zero-form moduli inserted at the center, the boundary gravitons would in effect be topological and activate only one independent coupling, the normalization of the on-shell closed $p$-form; for related discussions of ensembles in generalized Hamiltonian field theory, see \cite{Boulanger:2012bj}.
\item \emph{Three-dimensional models:} Besides the richness of two-dimensional 
conformal field theory, three-dimensional higher spin gravities, and in particular the 
Prokushkin--Vasiliev systems, provide a laboratory for further examination of several of 
the above issues, such as amplitudes involving solitons, the sewing operation, and the 
interplay between boundary and bulk gravitons, \emph{i.e.} the moduli contained in 
gauge functions and the Weyl zero-form, respectively. 
\item \emph{Tensionless strings and brane partons:}
Some of the motivation behind \cite{Colombo:2010fu} goes back to the work in \cite{Engquist:2005yt} on tensionless limits of strings and branes in anti-de Sitter spacetime,
where it was observed a negative cosmological constant, $\L<0$, plays 
a crucial role for the cusps which form on rotating strings and branes to 
behave as worldvolume solitons with carrying the spacetime quantum numbers of singletons.
It was then argued that these fundamental degrees, referred to as brane partons, are reincarnated in a number of dual pictures: i) as free quanta of conformal field theories on large $p$-branes in $AdS_{p+2}$; ii) as physical states in gauged quantum-mechanical models obtained by discretizing branes and then sending both tension to zero and $|\L|$ to infinity (in units of the lattice spacing);  iii) as vertex operators on boundaries of topological open strings in symplectic manifolds introduced in order to quantize (ii) covariantly; and iv) as deeper twist fields in topological Wess--Zumino--Witten models with critical $W_\infty$-algebras arising in continuum limits where multiple open strings are stacked on top of each other. 

In particular, some quantitative evidence was found supporting the idea that Vasiliev's equations arise from demanding consistency of disc-shaped topological open strings deformed by insertions of $\widehat A$ along their boundaries and $
\widehat \Phi$ into their bulks. This led the 
authors of \cite{Engquist:2005yt} to propose the identification of the leading order of the 
zero-form charges with, on the one hand, open-string amplitudes in the unperturbed 
background, and, on the other hand, dual twistor-space amplitudes for Vasiliev's theory. 
We thus view the results of the present paper as lending further support to topological 
open string picture proposed in \cite{Engquist:2005yt}.
It would be interesting to seek a more direct relationship between the twistor space action in \eqref{action} and the partition function of the topological open string, which might provide a concrete approach to constructing the twistor space regulator obeying Assumption (iii) in Section \ref{Sec:action}.

\item \emph{Hypercone and dual flow equations:} Another idea brought forth in 
\cite{Engquist:2005yt}, motivated by the need to send the cosmological constant to minus infinity in the tensionless limit, and that touches upon the holographic proposal \eq{holoprop}, is that the flow equations resulting from the perturbative expansion of Vasiliev's equations on the Dirac hypecone (rather than anti-de Sitter spacetime), which are ultra-local on the conifold, are holographically dual to the renormalization group equations governing the conformal field theory with 
finite sources, which in particular may be of relevance for \cite{Douglas:2010rc}.
\item \emph{Zero-form charges in gauged supergravities}: The group-theoretic reason 
underlying the existence of zero-form charges is that if the cosmological constant is non-
zero then the Weyl zero-form of a spin-$s$ field belongs to a nonlinear module $R_{(s)}$ of $so(2,3)$ containing a Lorentz covariant representation space, referred to as the spin-
$s$ twisted-adjoint representation, $T_{(s)}$ say, that is isomorphic to 
its dual, that is, $T_{(s)}\cong (T_{(s)})^\ast$; for further discussions, see \cite{Boulanger:2008up}. In other words, there exists a symmetric bi-linear form on $T_{(s)}
\otimes T_{(s)}$, which one can identify as the restriction of $I_2^{(2)}$ to the spin-$s$. 
More generally, the leading orders $I_{n_0}^{(n_0)}$ 
define $n_0$-linear symmetric forms on $T_{(s)}$, viewed as a linear representation space. 
Thus restricting to the case of fields with spins $s\leqslant 2$, one may ask whether these symmetric forms admit deformations by sub-leading terms as to remain closed on shell for nonlinear models, such as gauged supergravities, in which case their 
twistor-space realization would provide an intrinsic (re-)formulation of gravity in twistor 
space. 
\end{itemize}

\subsection{Summary}

\paragraph{Our Results}

Following the dual twistor-space method proposed in \cite{Colombo:2010fu} for obtaining manifestly gauge-invariant and scheme independent classical observables in four-dimensional higher spin gravity from zero-form charges, and employing a refined regularization scheme, we have obtained a number of quasi-amplitudes in the twistor space plane wave basis of which we have shown that a subset reproduces the point-split two- and three-current correlation functions in three-dimensional theories of free conformal scalars and fermions once their external legs are saturated with unfolded bulk-to-boundary propagators.

\paragraph{Proposal for Free Energy}

Thus, in the sector of twistor plane waves, we have found the following three types of zero-form charges (see also \cite{Sezgin:2011hq}):
\bea {\cal I}_{n_0}&=&{\rm Tr}\left[ (\widehat\Phi'\star\widehat {\bar\kappa})^{\star n_0}\star\widehat\kappa\widehat{\bar\kappa}\right]\ ,\quad n_0~=~2,4,\dots\ ,\\
{\cal I}'_{n_0}(\l)&=&{\rm Tr}\left[ (\widehat\Phi'\star\widehat {\bar\kappa})^{n_0}\star e_\star^{i\l\widehat S'} \right]\ ,\quad n_0~=~1,3,\dots\ ,\\
{\cal I}'_{n_0}(\l,\bar\l)&=&{\rm Tr}\left[ (\widehat\Phi'\star\widehat {\bar\kappa})^{n_0}\star e_\star^{i(\l\widehat S'-\bar\l\widehat{\bar S}')} \right]\ ,\quad n_0~=~2,4,\dots\ .\eea
As for ${\cal I}_{n_0}$, arguments in favor of it being finite and protected to all orders, \emph{i.e.} ${\cal I}_{n_0}^{(n)}=0$ for $n\geqslant n_0+1$, were given in \cite{Colombo:2010fu}, and its off-shell resolution as a topological vertex operator, \emph{viz.} 
\be {\cal I}_{n_0}~\approx~  -{\widehat Tr}'\left[4 n_0 \widehat F\star \widehat F(\widehat \Phi\star \widehat\kappa)^{n_0-2}+\frac{n_0-2}2 d^4Z  (\widehat \Phi\star \widehat\kappa) ^{\star(n_0-2)}\star \widehat\kappa\widehat{\bar\kappa}\right]\ ,\ee
was given in \cite{Sezgin:2011hq}. 
On the other hand, suitably smeared ${\cal I}'_{n_0}|_{\rm reg}=\int \frac{d^2\l d^2\bar\l}{(2\pi)^2} \widetilde{\cal V}_{n_0}(\l,\bar\l){\cal I}'_{n_0}(\l,\bar\l)$, where $\widetilde{\cal V}_{n_0}(\l,\bar\l)=2\pi \delta^2(\bar\l)  \widetilde{\cal V}_{n_0}(\l)$ for odd $n_0$, have finite perturbative expansions with non-trivial sub-leading corrections ${\cal I}_{n_0}^{(n)}$, weighted by couplings given by Taylor coefficients and moments of the regularization functions.
Drawing on analogies with the Wilson loop, we have proposed that the regularization functions can be determined uniquely by demanding the existence of an off-shell resolution as a topological vertex operator (and that the resulting couplings are finite on shell).
Under this assumption, one can define the twistor space action \eqref{action}, 
and fix normalizations by demanding the the point-split amplitudes with bulk-to-boundary propagators reproduce the canonically normalized correlation functions; the latter analysis can be performed once and for all at a fixed value of $b$, and can hence be simplified by taking $b^2=\pm 1$. 

To recapitulate, under the arguably mild assumptions summarized in the beginning of  Section \ref{Sec:action}, we have proposed a dual twistor-space method for obtaining the full on-shell actions for four-dimensional higher spin gravities of Vasiliev type with off-shell formulations based on generalized Hamiltonian actions 
 \cite{Boulanger:2011dd} (see also \cite{Sezgin:2011hq,Boulanger:2012bj}) as follows:
\begin{itemize}
\item \emph{On-shell procedure (extracting quasi-amplitudes and quasi-correlators:)}
\begin{itemize}
\item[] \emph{Step ia)} The zero-form charges are expanded perturbatively using the regularization scheme spelled out in Section \ref{Sec:reg} as to obtain quasi-amplitudes for twistor space plane waves; 
\item[] \emph{Step ib)} Unfolded bulk-to-boundary propagators are attached to the external legs using Gaussian integration; 
\item[] \emph{Step ic)} The resulting conformal quasi-correlation functions are decomposed into point-split and point-contact contributions (of which the former may appear at leading as well as sub-leading orders);
\end{itemize}
\item[] At this stage, the free parameters in the on-shell action are given by the chemical potentials $\mu_{n_0}$ and couplings given by moments of the regularization functions $\widetilde{\cal V}_{n_0}$ and its Fourier transform.
\item \emph{Off-shell procedure (fixing normalizations)\footnote{Step (iia) can be carried out at the level of the full master fields while Step (iib) is perturbative.}:}
\begin{itemize}
\item[] \emph{Step iia)} Regularization functions are fixed up to overall normalizations by demanding that ${\cal I}'_{n_0}$ are the on-shell values of topological vertex operators;
\item[] \emph{Step iib)} Overall normalizations are fixed recursively by demanding cluster decomposition of point-split correlation functions.
\end{itemize} 
\item[] The end result is the effective action with free parameters given by the chemical potentials $\mu_{n_0}$.
\end{itemize}

\paragraph{Outlook}

Although it appears to us that a number of interesting directions are now free to explore, we would nonetheless like to stress the fact that, at present, we cannot claim the existence of any novel quasi-amplitudes, corresponding holographically to anything else than point-split multi-current correlation functions in free theories, which are in some sense trivial.
Thus, the highest priority is clearly to establish whether such structures arise or not. 
It would also be interesting to identify parity violating contributions to holographic correlation functions needed for the conjectures involving self-interacting conformal field theories made in \cite{Giombi:2011kc,Aharony:2011jz}; we expect that the simplest such function, namely the parity violating three-point function \cite{Giombi:2011rz}, arises in the second sub-leading order, namely in ${\cal I}^{\prime(3)}_{1}$, which in principle may depend on the parity-breaking parameter $\theta_0$ in a nontrivial fashion. 

Assuming a nontrivial outcome, we would claim that the perturbative expansion of the Vasiliev system around anti-de Sitter spacetime contains physical information in addition to that contained in point-split multi-current correlation functions in free theories, namely the data required to construct the fully nonlinear generating function for conformal field theories deformed by finite sources.
It is in this spirit, and by exploiting recent progress in the off-shell formulation of higher spin gravity, that we have proposed a bulk counterpart given by a sum of normalized zero-form charges. 
We expect this quantity to be useful both at the classical level, where it can be interpreted as a regularized free energy free for configurations that are asymptotic to anti-de Sitter spacetime, free from gauge artifacts and other scheme dependencies, 
and at the quantum level, where it may function as the evolution kernel for master fields in twistor space.

\paragraph{ Acknowledgements:} We are grateful to N. Boulanger, C. Iazeolla and E. Sezgin for communications and insightful remarks during the course of the project. We would also like to thank I. Bandos, A. Castro, D. Chialva, T. Damour, V. Didenko, S. Giombi, P. Kraus, D. Ponomarev, A. Sagnotti, E. Skvortsov, D. Sorokin, Ph. Spindel, M. Valenzuela, M. Vasiliev and Xi Yin for discussions. P.S. would like to acknowledge the Department of Mathematics and the ICBM at the Bosphorus University for hospitality during the finalization of the work.

\paragraph{Note added:} Our proposal in Eqs. \eq{propamodel} and \eq{propbmodel} has been verified at the level of totally symmetric structures in \cite{Didenko:2012tv}.

\begin{appendix}
\section{Expansion of $e_{\star}^{i \m \widehat S} $}\label{Exponential}

The weak-fields expansion of the regularized zero-form charges given in \eqref{reg}  involves the $\Phi^{\prime }$ expansion of the $\star$-exponentials $e_{\star}^{i \m \widehat S'} $.
From the definition of $\star$-exponentials one has 
\bea
\label{star exponential}
e_*^{z + \widehat V'} = 1 + \left(z + \widehat V'\right) + \frac{1}{2!} \left( (z + \widehat V') \star  (z + \widehat V') \right) + \frac{1}{3!} \left( (z + \widehat V') \star  (z + \widehat V') \star (z + \widehat V') \right) \dots
\eea
where we have defined
\be
z := i \m^{\a} z_{\a} \ ,  \quad  \widehat V' := 2  \m^{\a}  \widehat V'_{\a}\ ,  \quad  \bar z := i \mb^{\a} \zb_{\ad} \ \ . 
\ee
The perturbative expansion of \eqref{star exponential} is obtained by considering that in the 0-order $A^{(0)}_{\a} = 0$ implies  
\be
 \left( e_*^{z + \widehat V'} \right)^{(0)} = 1 + z + \frac{1}{2!} (z)^{*2}  + \frac{1}{3!} (z)^{*3} + \dots =  e_*^z \ ,
 \ee
and at the first sub-leading order  
\bea
 \left( e_*^{z + \widehat V'} \right)^{(1)}  &=&  \widehat V'^{(1)} + \frac{1}{2!} \left(  z \star \widehat V'^{(1)}  + \widehat V'^{(1)}  \star z \right) + \frac{1}{3!} \left( z \star z\star  \widehat V'^{(1)}  + z \star \widehat V'^{(1)} \star z + \widehat V'^{(1)} \star z \star z \right) + \dots   \\ [5pt]
  &=&  \widehat V'^{(1)} + \frac{1}{2!} \left(  2  z \star \widehat V'^{(1)}  + [ \widehat V'^{(1)}, z ]_* \right) + \frac{1}{3!} \left(3  z \star z \star \widehat V'^{(1)}  + 3 z \star  [ \widehat V'^{(1)}, z ]_*  +   \left[ [ \widehat V'^{(1)}, z]_* , z \right]_*  \right) + \dots \nonumber \\ [5pt]
  &= &   \widehat V'^{(1)} + \frac{1}{2!} \left( 2 z  + (2i ) \partial_z \right) \star \widehat V'^{(1)} + \frac{1}{3!} \left( 3 z \star z   +  (2i) 3 z \star \partial_z + (2i)^2 \partial_z\partial_z  \right) \star  \widehat V'^{(1)} + \dots  \ , \nonumber
   \eea
where we use the $\star$-product relation $ [f, z] = (2 i) \partial_z f $, with the definition $ \partial_z :=  i \m^{\a} \partial_{\a}$\footnote{We have also rearranged terms as follows  
\bea
 \widehat V'^{(1)} \star (z)^{*N}  &=&  (z)^{* N} \star  \widehat V'^{(1)}~ +~   N \  (z)^{* n -1} \star  [ \widehat V'^{(1)}, z ]  ~+ ~ (N-1) (z)^{* N -2} \star  [ \widehat V'^{(1)}, z ]  , z] + \nonumber \\[5pt] 
 &&\qquad+ \dots +  [ [\dots [[ \widehat V'^{(1)} , z], z] \dots ] ,z ] 
\eea}.  Factorizing an overall factor $e_*^z$ one obtains\footnote{We perform the infinite sum using 
\be
\sum_{n=0}^{\infty} \frac{x^n}{(n+1)!} = \sum_{n=0}^{\infty} \frac{x^{n+1}}{(n+1)!} \frac{1}{x} =\frac{e^{x}- 1}{x} \ . 
\ee }
 \be
 \left( e_*^{z + \widehat V'} \right)^{(1)} ~=~  e_*^z \star  \left(  \widehat V'^{(1)}+  \frac{(2i)}{2!} \partial_z   \widehat V'^{(1)} +  \frac{(2i)^2}{3!} \partial^2_z   \widehat V'^{(1)}  + \dots \right)  ~=~ e_�^z \star \left( \frac{e^{ (2i) \partial_z} - 1}{(2i)\partial_z} \right) \widehat V'^{(1)}
 \ee
Similar expressions hold for the complex conjugate. Then we can substitute the plane wave expansion of the first order master field  \eqref{Auno} and obtain
\bea
\left( e^{i \m^{\a} \widehat S'_{\a}} \right)^{(1)} &=&   - i b \ \m z \int^{1}_{0} dt t \  \left(  \frac{e^{ 2 i \m (y - \l ) t } - 1}{2 i \m (y - \l ) t  } \right) e^{i (y - \l  - \m (1 - \frac{1}{t}))( z + \m) t  \  - i \lb  \yb } \\ [5 pt]
\left( e^{i \mb^{\ad } \widehat {S'_{\ad}}} \right)^{(1)} &=&   - i \bar b \ \mb \zb \int^{1}_{0} dt  t \  \left( \frac{e^{ - 2 i \mb (\yb + \lb ) t } - 1}{- 2 i \mb (\yb + \lb ) t  } \right) e^{i  \l y  \ - i (\yb + \lb - \mb (1 - \frac{1}{t})) ( \zb + \mb)t } 
\eea
where we have used \eqref{emuzstarf1} and \eqref{emuzstarf2}.

\section{Note on the second order integrals}
We can now proceed to the explicit evaluation of the various second order terms. 
Depending on which master field is corrected one obtains two similar structures \emph{i.e.}
\be
\label{second order structures}
\widehat{\rm Tr}\left[ f(Y,Z,\L) \int_{0}^{1} dt_1 dt_2   t_1  e^{ i A t_1 t_2  + i B (1 - t_1)} \right] \ ,  \quad  \widehat{\rm Tr}\left[ g(Y,Z,\L)   \int_{0}^{1} dt t \left( \frac{ e^{i C t} -1}{i C t}   \right)e^{ i D  t}  \right] 
\ee
that correspond respectively to the expansion of the master 0-form $\widehat \Phi$ and the functions $f(\widehat S, \widehat {\bar S} )$.
The factors $A,B,C,D$ that are linear combinations of spinors variables $Y,Z, \L_i$, $\L_i$ being linear combinations of the external momenta and we have also redefined the twistor variables $Y, Z$ for simplicity .
The functions  $ f = f(Y,Z,\L) $, $g =  g(Y,Z,\L)  $ do not depend on the $t_i$-parameters and are regular function in all their variables.    
The explicit expression of the factors $A,B,C,D$ and  the functions $ f = f(Y,Z,\L) $, $g =  g(Y,Z,\L)  $ are obtained by performing all the $\star$-product appearing in some $\widetilde{\cal I}^{(n)'}(\Phi_1', \dots, \Phi'_n)$. 
The homotopy integration is trivial and gives
\bea
\label{schematic form 1}
\widehat{\rm Tr}\left[ f(Y,Z,\L) \int_{0}^{1} dt_1 dt_2   t_1  e^{ i A t_1 t_2  + i B (1 - t_1)} \right]  ~=~  \widehat{\rm Tr}\left[  (-)f(Y,Z,\L)  \left( \frac{e^{iA } - e^{iB}}{A (A-B)} + \frac{e^{iB}-1}{A B}\right) \right]
\eea
or
\bea
\label{schematic form 2}
\widehat{\rm Tr}\left[ g(Y,Z,\L)   \int_{0}^{1} dt t \left( \frac{ e^{i C t} -1}{i C t}   \right)e^{ i D  t}  \right]  ~=~  \widehat{\rm Tr}\left[   (-)g(Y,Z,\L)  \left( \frac{e^{i(C+ D) } - 1}{ C (C+D) } - \frac{e^{iD} - 1}{CD} \right) \right] \ .
\eea
The two ratios of $A(Y,Z),B(Y,Z)$ under round brackets in \eqref{schematic form 1}  are not singular in the limits $A \to B$ or $A \to 0 $ and $B \to  0$.

\section{Cauhy Principal Value}
We want to compute the integral
\be
I ~=~ \int_{ - \infty}^{\infty} \frac{f(x)}{x}
\ee
where $f(z)$ is analytic in the upper/lower  half plane and $|f(z)| \to 0 $ for $|z| \to \infty $. One consider the complex integral defined by
\be
J ~=~ \int_{C} \frac{f(z)}{z} ~  = ~  \int_{\Gamma_1}dz \frac{f(z)}{z} ~+~  \int_{\Gamma_2} dz \frac{f(z)}{z}~+~  \int_{\Gamma_3}dz  \frac{f(z)}{z} ~=~ J_1~+~ J_2~+~ J_3 
\ee
where 
\bea
\Gamma_1 & : &  z  = x; \qquad  \{  - R \le  x \le - \e \} U \{\e \le  x \le R  \} \\[5pt]
\Gamma_2 &:  &  z  = \e e^{i\theta} ; \qquad   \pi \ge\theta \ge  0 \ ;   \\[5pt]
\Gamma_3 &: &   z  = R e^{i\theta} ; \qquad    0 \le \theta \le \pi \ .
\eea
Since $|f(z)| \to 0 $ for $|z| \to \infty $,  in the limit $\e \to   0 $ and $R \to \infty$ the integral becomes
\bea
J ~ = ~  I  - \lim_{\e \to 0} \int_{0}^{\pi} d\theta  \e e^{i \theta}  \frac{f(\e e^{i\theta} )}{ \e e^{i\theta}}~+~ 0  = I - i \pi f(0) ~=~ 0 
\eea 
by applying the definition of Cauchy Principal Value ($J_1 \to I  $) and the theorem of residues ($J$  =  0 ). 

\section{Details of calculation of two-point functions}\label{App:Ampl}

Upon splitting 
\be \widetilde \Phi'_i~\equiv~ \sum_{\s_i , \bar \s_i =\pm 1}b(B_i+\overline B_i)\ ,\quad B_i~=~\frac{1}{2}K_i e^{i y \boldsymbol{\Sigma}_i \yb+i \bar \s_i \nb_i \boldsymbol{\Sigma}_i y}\ ,\quad   \overline B_i~=~\frac{b^2}{2}K_i e^{i y \boldsymbol{\Sigma}_i \yb+i  \s_i \nu_i \boldsymbol{\Sigma}_i \yb }\ ,\label{SplitPhi}\ee
we can expand \eqref{amplitude 2} and \eqref{amplitude 2'} as 
\bea
{\cal I}_2^{(2)'}(\vec x_1,\chi_1; \vec x_2,\chi_2) &  = & \int  \frac{d^4\L_1}{(2 \pi)^2} \frac{d^4\L_2}{{(2 \pi)^2}} \  \widetilde \Phi'_1(r_0,\vec x_0;\vec x_{1},\chi_1 ; \L_1)  \widetilde \Phi'_2(r_0,\vec x_0;\vec x_{2},\chi_2; \L_2)    {\cal I}^{(2)'}_{2} (\L_1, \L_2)  \nonumber \\ [5pt]   
& \equiv &  \frac{b^2}{4} K_{1}K_{2}  \sum_{\s_1, \s_2 } \sum_{ \bar \s_1 , \bar \s_2 }  \left( 1 2 + \bar 1 2  +  1 \bar 2 +  \bar 1 \bar 2 \right) \ ,  
\eea
\bea
{\cal I}^{\prime(2)'}_2(\vec x_1,\chi_1; \vec x_2,\chi_2 ) &  = & \int  \frac{d^4\L_1}{(2 \pi)^2} \frac{d^4\L_2}{{(2 \pi)^2}} \  \tilde \Phi_1(r_0,\vec x_0;\vec x_{1},\chi_1; \L_1)  \tilde \Phi_{2}(r_0,\vec x_0;\vec x_{2},\chi_2; \L_2)     {\cal I'}^{(2)'}_{2} (\L_1, \L_2)  \nonumber \\ [5pt]   
& \equiv & \frac{b^2}{4} \widetilde {\cal V}'_2(0,0)  K_{1}K_{2}   \sum_{\s_1, \s_2 } \sum_{ \bar \s_1 , \bar \s_2 }  \left( (1 2)' + (\bar 1 2)'  +  (1 \bar 2)' +  (\bar 1 \bar 2)' \right)   \ ,  
\eea
where the separate contributions are given by
\bea
(1 2)  & = &  \int \frac{d^2 \l_1d^2 \l_2 d^2\lb_1 d^2\lb_2 } {(2 \pi)^{4} } e^{ - i    \l_1 \boldsymbol{\Sigma}_1 \lb_1  -  i   \l_2 \boldsymbol{\Sigma}_2 \lb_2} e^{ i   \s_1 \l_1 \nu_1 + i   \s_2 \l_2 \nu_2} \frac12 \sum_{{\rm perm.}\,1,2} e^{ - i \l_1 \l_2  -  i \lb_1 \lb_1 }    \nonumber\\ [5pt]
& = &  \sum_{{\rm perm.}\,1,2}\int\frac{ d^2 \l_1 d^2\l_2}{(2 \pi)^2}  e^{ -  i \l_1 (1  - \boldsymbol{\Sigma}_1 \boldsymbol{\Sigma}_2) \l_2}  e^{ i   \s_1 \l_1 \nu_1 + i   \s_2 \l_2 \nu_2}   \nonumber \\[5pt]
& = &  \sum_{{\rm perm.}\,1,2}\frac{1}{\det{(1  - \boldsymbol{\Sigma}_1  \boldsymbol{\Sigma}_2)} } e^{  - i  \s_1 \s_2 \nu_1  \frac{ 1  - \boldsymbol{\Sigma}_1  \boldsymbol{\Sigma}_2  }{\det{(1  - \boldsymbol{\Sigma}_1  \boldsymbol{\Sigma}_2)}  } \nu_2 }  \nonumber \\[5pt]
 & = &   \frac{1}{4}\frac{1}{K_1 K_2}\frac{1}{\vec{x}_{12}^2  } \sum_{{\rm perm.}\,1,2} e^{   i \s_1 \s_2  P_{12} } \  ,  \\[5pt] 
(\bar 1 2)   &=&  ( 1 \bar 2)  ~=~ b^2  \frac{1}{4}\frac{1}{K_1 K_2}\frac{1}{\vec{x}_{12}^2 } \ ,  \\[5pt]
 (\bar 1 \bar 2)  &= & (1 2)  \ , \\[5pt]
(1 2 )'  &=& (\bar 1 \bar  2 )'  ~=~ (\bar 1 2) \ ,  \\[5pt]
(\bar 1 2)' &=& (1 \bar 2)' ~=~  (1 2)    \ ,  
\eea
using the identities \eqref{sigma identities} for the $\boldsymbol{\Sigma}_i$-matrices and the definition of $P_{12}$ given in \eqref{defP12}. 
Thus one has 
\bea
{\cal I}_2^{(2)'}(\vec x_1,\chi_1; \vec x_2,\chi_2) &  = &  \frac{1}{ 16  \vec{x}_{12}^2  } \sum_{{\rm perm.}\, 1, 2} ~\sum_{\s_1, \s_2 }~ \sum_{ \bar \s_1 , \bar \s_2 }  ((12)   +   (\bar 1 2) ) \nonumber \\[5pt]
&  = &  \frac{2}{ \vec{x}_{12}^2  }   \left(  1 +b^2 \cos \left( P_{12}\right) \right)  
 \eea
\bea
{\cal I}^{\prime(2)'}_2(\vec x_1,\chi_1; \vec x_2,\chi_2 )  &  = &  b^2  \widetilde {\cal V}'_2(0,0) \frac{1}{ 16 \vec{x}_{12}^2  } \sum_{{\rm perm.}\, 1, 2} ~\sum_{\s_1, \s_2 } ~\sum_{ \bar \s_1 , \bar \s_2 }  ((12)   +  (\bar 1 2) ) \nonumber \\[5pt]
&  = & \widetilde {\cal V}'_2(0,0) \frac{2}{ \vec{x}_{12}^2  }   \left(  b^2+\cos \left( P_{12}\right) \right)   
\eea

\end{appendix}

\bibliographystyle{utphys}

\providecommand{\href}[2]{#2}\begingroup\raggedright\endgroup

\end{document}